\documentclass[sigconf, authorversion, nonacm]{acmart}

\AtBeginDocument{%
  }

\usepackage{xspace}
\usepackage{listings}
\usepackage{multirow}

\usepackage{epigraph}
\usepackage{tikz}

\usepackage{cleveref}
\usepackage[inline]{enumitem}

\usepackage{colortbl}
\usepackage{booktabs}
\usepackage{array}
\usepackage{rotating}

\definecolor{CCircle}{HTML}{000000}

\newcommand{\ccircle}[1]{%
\begin{tikzpicture}[enum/.style={circle,draw=black,very
    thin,fill=CCircle,text=white,inner sep=1pt},baseline=-3pt]
  \node (node 1) at (0,0) [enum] {\scriptsize #1};%
\end{tikzpicture}}

\usepackage[T1]{fontenc}
\usepackage{listings}

\lstdefinelanguage{TSX}{
  language=Java,
  morekeywords={
    typeof,new,true,false,catch,function,return,null,switch,var,let,const,
    if,in,while,do,else,case,break,class,export,boolean,throw,implements,
    import,this,extends,interface,type,keyof,readonly
  },
  sensitive=true,
  morestring=[b]",
  morestring=[b]',
  morecomment=[l]{//},
  morecomment=[s]{/*}{*/},
  moredelim=[s][\ttfamily]{`}{`},
}

\lstdefinestyle{code}{
  basicstyle=\ttfamily\small,
  keywordstyle=\color{blue}\bfseries,
  stringstyle=\color{red},
  commentstyle=\color{gray},
  showstringspaces=false,
  columns=fullflexible,
  upquote=true,
  mathescape=false,
  breaklines=true
}

\newcommand{\say}[1]{``#1''}

\newcommand{\tool}{\textsc{Celestial}\xspace}
\newcommand{\alex}{Chloe\xspace}
\newcommand{\sbook}{Storybook.js\xspace}

\newcommand{\smalltt}[1]{%
  {\small\ttfamily
   \hyphenchar\font=`\-\relax
   #1}%
}

\newcommand{\componentAnalyzer}[0]{\smalltt{ComponentAnalyzer}}
\newcommand{\naturalSampler}[0]{\smalltt{MimeticSampler}\xspace}
\newcommand{\coverageAnalyzer}[0]{\smalltt{CoverageAnalyzer}\xspace}
\newcommand{\midsepremove}{\aboverulesep = 0mm \belowrulesep = 0mm}

\begin{document}

\title[Instantiating UI Components with Distinguishing Variations]{The Way We Notice, That's What Really Matters: Instantiating~UI~Components with Distinguishing Variations}

\settopmatter{authorsperrow=4}

\author{Priyan Vaithilingam}
\orcid{0000-0001-6730-5683}
\affiliation{%
  \institution{Apple}
  \city{Seattle}
  \state{Washington}
  \country{USA}
}
\email{priyan@apple.com}

\author{Alan Leung}
\affiliation{%
  \institution{Apple}
  \city{Seattle}
  \state{Washington}
  \country{USA}}
\email{alleu@apple.com}

\author{Jeffrey Nichols}
\affiliation{%
  \institution{Apple}
  \city{Seattle}
  \state{Washington}
  \country{USA}
}
\email{jwnichols@apple.com}

\author{Titus Barik}
\affiliation{%
  \institution{Apple}
  \city{Seattle}
  \state{Washington}
  \country{USA}
}
\email{tbarik@apple.com}

\renewcommand{\shortauthors}{Vaithilingam et al.}

\begin{abstract}

Front-end developers author UI components to be broadly reusable by parameterizing visual and behavioral properties. While flexible, this makes instantiation harder, as developers must reason about numerous property values and interactions. In practice, they must explore the component’s large design space and provide realistic and natural values to properties. To address this, we introduce \emph{distinguishing variations}: variations that are both mimetic and distinct. We frame distinguishing variation generation as design-space sampling, combining symbolic inference to identify visually important properties with an LLM-driven mimetic sampler to produce realistic instantiations from its world knowledge.

We instantiate distinguishing variations in \tool, a tool that helps developers explore and visualize distinguishing variations. In a study with front-end developers ($n=12$), participants found these variations useful for comparing and mapping component design spaces, reported that mimetic instantiations were domain-relevant, and validated that \tool transformed component instantiation from a manual process into a structured, exploratory activity.
\end{abstract}

\begin{teaserfigure}
  \centering
  \includegraphics[width=0.9\linewidth]{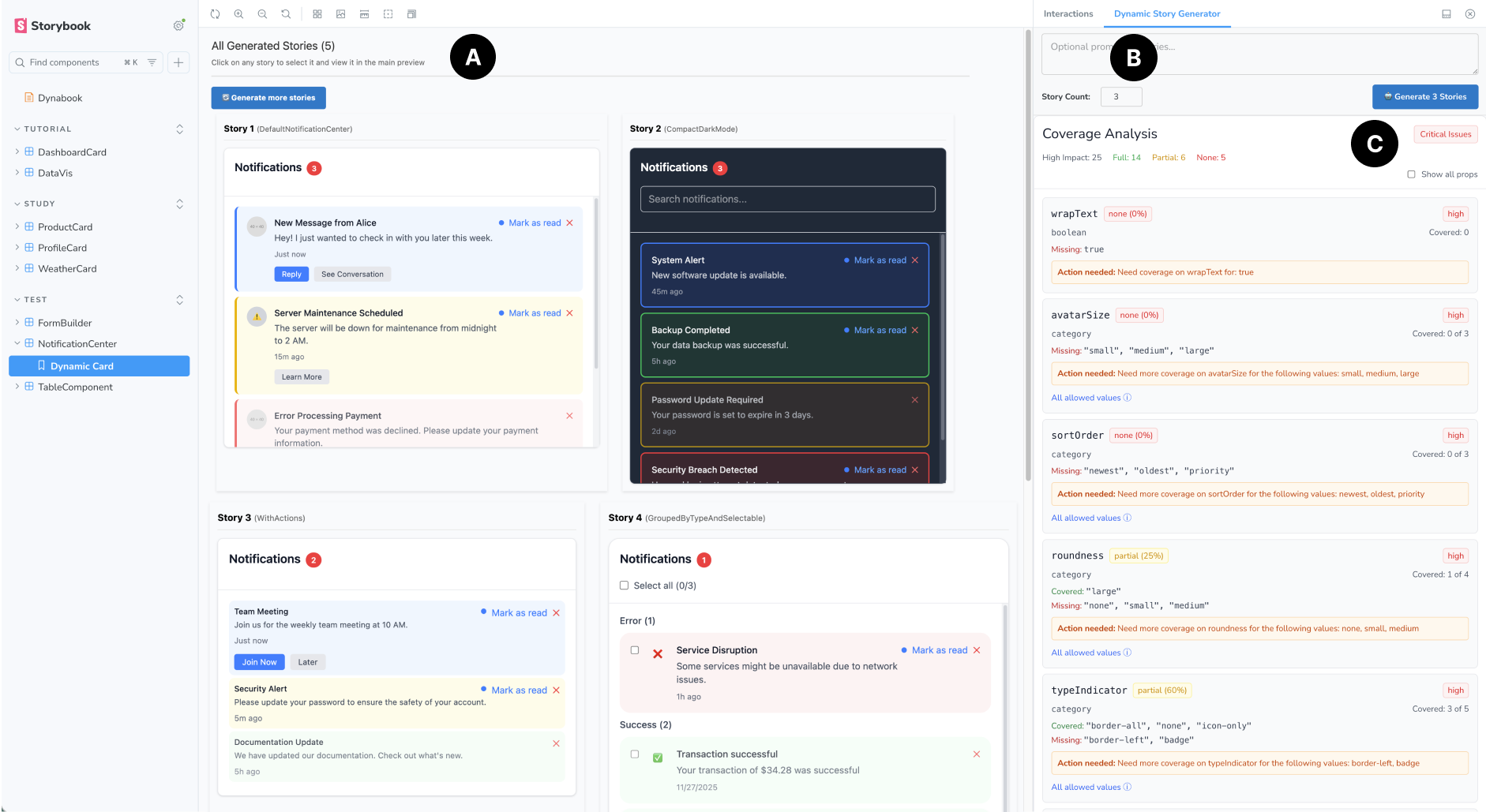}
  \caption{The \tool user experience. \textmd{\protect\ccircle{A} Explore distinguishing variations that are distinct and mimetic, spanning the entire design space of the UI components. \protect\ccircle{B} Generate additional variations to explore by providing an optional grounding instruction. \protect\ccircle{C} Obtain a comprehensive coverage analysis over the design space for the generated variations.}}
  \Description{This image shows a screenshot of the \tool user experience. In the center pane, the developer can browse distinguishing variations that are distinct and mimetic, spanning the whole design space of the UI component. At the top of the right pane, users can give optional instructions to ground the sampling for further variations that they want to explore. Finally, below the prompt box, the users can get a comprehensive coverage analysis over the design space for the generated variations.}
  \label{fig:teaser}
\end{teaserfigure}

\maketitle


\newpage

\section{Introduction}





\vspace{6pt}
\begin{tabular}{l}
We were made to be nothing more than this\\
Finding magic in all the smallest things\\
The way we notice, that's what really matters\\
Let's make tonight go on and on and on\\
\hline
\qquad\qquad\qquad Ed Sheeran, \emph{Celestial} (2022)
\end{tabular}\vspace{6pt}

\begin{figure*}[h]
    \centering
    \includegraphics[width=0.9\linewidth]{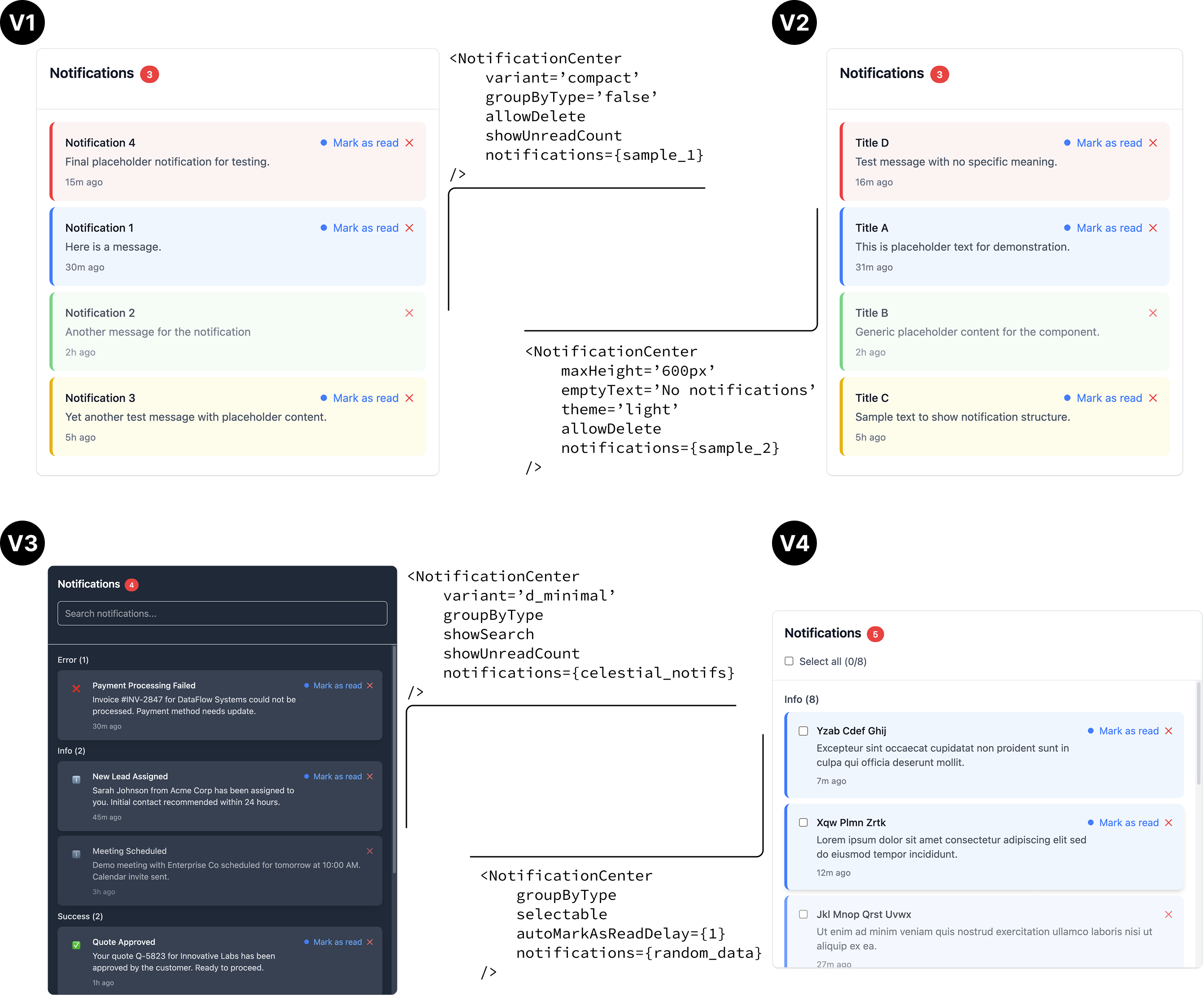}
    \caption{Four different valid instantiations of \texttt{NotificationCenter} component. \textmd{Although the code to instantiate the variations only differ slightly from each other, visually \protect\ccircle{V1} and \protect\ccircle{V2} look very similar, while \protect\ccircle{V1} and \protect\ccircle{V3} look substantially different. While \protect\ccircle{V4} is a valid instantiation, it looks \emph{unnatural}, violating real-world conventions.}}
    \Description{This image shows screenshots of four different variations of the weather component (V1 - V4). Although the code to instantiate the variations only differs slightly from each other, visually, V1 and V2 look very similar, while V1 and V3 look substantially different. While V4 is a valid instantiation, it looks unnatural, violating real-world conventions (e.g., location and temperature scale).}
    \label{fig:variations}
\end{figure*}

Consider \alex, a front-end developer integrating a notification center into her company dashboard. To do so, she must first understand how to use the \smalltt{NotificationCenter} component from an internal UI component library.  Since the component is designed to support a diverse set of requirements from different teams, its behavior depends heavily on its properties and data used to instantiate it. However, the documentation only contains a few abstract examples with some placeholder data, and is not often up to date due to constantly changing requirements and code. As a result, to understand the component behavior, \alex is forced to manually instantiate additional variations that can show the extent of the design space with data specific to her use-case. However, the dozens of properties and nested configurations that make \smalltt{NotificationCenter} expressive also create a design space of thousands of possible variations. Here are just a few ways \alex can instantiate this component (see these rendered in~\Cref{fig:variations}):

\begin{description}
    \item[V1:]
\begin{lstlisting}[language=TSX,style=code]
<NotificationCenter variant='compact' 
    groupByType='false' showUnreadCount 
    allowDelete notifications={sample_notif1} />
\end{lstlisting}

    \item[V2:]
\begin{lstlisting}[language=TSX,style=code]
<NotificationCenter maxHeight='600px'
    emptyText='no notifications' theme='light'
    allowDelete notifications={sample_notif2} />
\end{lstlisting}

    \item[V3:]
\begin{lstlisting}[language=TSX,style=code]
<NotificationCenter variant='d_minimal'
    groupByType showSearch showUnreadCount 
    notifications={celestial_notifs} />
\end{lstlisting}

    \item[V4:]
\begin{lstlisting}[language=TSX,style=code]
<NotificationCenter autoMarkAsReadDelay='1'
    selectable groupByType 
    notifications={random_data} />
\end{lstlisting}
\end{description}

Learning to use this component presents several challenges:
(1) \alex must learn all of the component's properties to instantiate it. For complex components like \smalltt{NotificationCenter}, this can mean dozens of properties---greatly increasing the effort to get started. This effort is higher for nested compound components because they require constructing nested child data (e.g., a list of \smalltt{Notification} objects), each with its own properties and constraints, before the component can be meaningfully rendered.
(2) \alex needs to understand how these properties affect the visual layout and how they interact with each other. For example, although the property values differ between V1 and V2, the rendered output is almost identical. By contrast, small changes in property values between V1 and V3 produce substantially different renders.
(3) For complex and nested components like the \texttt{NotificationCenter}, \alex needs to provide realistic, domain-specific data for the component for her to fully understand its behavior---for example, how the notifications are rendered when there is media and whether it supports right-to-left languages like Arabic. Take V1, for instance: though the component is rendered correctly, the placeholder text shows no diversity.
(4) For \alex to explore the entire design space of the component, she will likely require instantiating more variations than just four. One way to automate this approach is to randomly populate property values based on the property type. However, doing this often produces unnatural and \emph{artificial} outputs as shown in V4.

When it comes to exploring UI component variations \emph{more is not necessarily better}~\cite{nagappan2013diversity, tiwari2020}. 
The challenge is not merely the size of the design space but that most possible variations are either unremarkable (do not change behavior) or implausible (break real-world conventions). To effectively understand the design space, we need to find the \emph{distinguishing} variations.

Let's look at the \texttt{NotificationCenter} again. If every variation the developer sees simply has a different notification title and message, the component may be technically varied but still functionally redundant. If instead one version enables \smalltt{showSearch}, but has only the placeholder ``Lorem Ipsum'' text across all the notifications, it is distinct but unnatural and not useful to explore the search feature. Consequently, a useful variation must first be \emph{mimetic}: the values mimic real-world conventions, so that each variation \emph{feels} like something that could appear in real-world use. Yet a collection of mimetic variations that all look the same with the same layout and colors still teaches the developer little about the component. To be valuable as a set, the examples must also be \emph{distinct}: each variation should surface a substantially different facet of the design space---such as comparing compact vs.\ detailed layouts---or exploring how storm warnings render in dark mode, and so on.

To address both requirements, we introduce \emph{distinguishing variations}, adapted from the principle of \emph{distinguishing inputs}~\cite{jha2010oracle, tiwari2020}. In program synthesis, distinguishing inputs should effectively differentiate two non-equivalent candidate programs. In our adaptation, \emph{distinguishing variations} should ensure every single instantiation of a component is mimetic; a collection of them must be distinct. Together, distinguishing variations allow developers to quickly grasp how a component behaves across realistic yet meaningfully diverse scenarios. 


To understand the challenges developers face when exploring UI components, we conducted a formative study with $9$ front-end developers. Our study revealed that examples (variations) in shared component libraries are often abstract, under-specified, and insufficient. The burden of instantiating and exploring distinguishing variations relevant to their scenario often falls on the developer who is consuming the UI components (\Cref{sec:formative}). 

We pose the task of generating distinguishing variations of UI components as a design-space sampling problem. We propose a hybrid approach that integrates symbolic methods---such as static analysis and type inference---to identify visually impactful properties of the component. These properties are then used by an LLM-powered \emph{mimetic sampler} that produces values and content that are distinct and mimetic. Together, this hybrid approach generates variations that surface expressiveness without clutter.

We instantiate this approach with our tool \tool, implemented as a \sbook~\cite{storybook} add-on. \sbook is a popular environment for developing and maintaining UI component libraries for the web. Within \sbook, \emph{stories} showcase representative variations of different UI components~\cite{storybookWhatsStory}, but these variations must be authored manually by the developer. We built \tool to automatically instantiate distinguishing UI component variations and help developers explore these variations within their familiar \sbook environment.
    
Through a user study with $12$ front-end developers, 
we found that distinguishing variations helped participants successfully discern and map the component design space, uncover serendipitous possibilities, and identify mimetic instantiations to be domain relevant. Participants also reported that \tool turned instantiating components from a manual, tedious task into a structured and exploratory activity.

In this paper, we make the following contributions: 

\begin{itemize}
\item We introduce a hybrid approach to generate distinguishing UI component variations that combines \emph{symbolic techniques} (e.g., static analysis and type inference) with a  \emph{mimetic sampler}.
\item We instantiate this approach in \tool, which allows developers to explore component variations directly within a \sbook environment. 
\item We present a study with professional front-end developers and found that distinguishing variations helps participants explore component design space while also producing results that were mimetic and distinct.
\end{itemize}


\section{A Demo of \tool}

Let's return to \alex from the Introduction, who is eager to integrate \texttt{NotificationCenter} from a shared enterprise component library into a customer relationship management (CRM) dashboard. Since this dashboard ships globally and surfaces high-stakes data (customer updates, approvals, outages), \alex needs to distinguish how the component behaves across meaningful variations. Hence, the examples should render mimetic representations of real-world CRM content, including media attachments and support for right-to-left (RTL) locales---not just placeholder text---before committing to use the component. She opens \sbook and navigates to the \texttt{NotificationCenter} page. A long list of properties immediately confronts her: layout controls such as \texttt{notificationStyle}, \texttt{typeIndicator}, \texttt{sortOrder}, etc., and styling parameters like \texttt{theme}, \texttt{size}, \texttt{backgroundStyle}, and \texttt{avatarStyle}. But these knobs are only the surface. Many outcomes depend on nested content and interactions between parent and child configuration: changing a \texttt{NotificationCenter} property can alter not only the outer container but also how individual \texttt{Notification} items are laid out, styled, and grouped. To meaningfully exercise the component, \alex must also instantiate the \texttt{Notification} objects it renders, providing structured content (sender, timestamps, actions, media, and metadata) whose internal dependencies determine whether a given variation is even plausible. Scrolling through the property list, she pauses: \say{Where do I even begin?} The handful of existing stories reuse the same flat placeholder data and cover only a thin slice of the design space, leaving key questions unanswered: \say{How does this render with images?} and \say{Does it support right-to-left languages like Arabic?}

\alex turns to \tool, available directly within \sbook under the \texttt{Dynamic Story} subsection for \texttt{NotificationCenter}. With a single click, \tool synthesizes a gallery of variations that span layout, theming, sorting, and grouping, while also automatically populating the nested \texttt{Notification} items with coherent, use-case-relevant content (\Cref{fig:teaser}~\protect\ccircle{A}). Unlike the existing mock-ups, these examples are mimetic: notification text aligns with a specific scenario, grouped categories match the message content, and the resulting gallery feels like something that could plausibly appear in the real world. While scanning the variations, \alex notices one that exposes a search feature. She had not considered search as part of the notification experience; seeing it in context prompts a new idea: she could let users filter notifications in-place instead of adding a separate dashboard-wide search.

\begin{figure}
    \centering
    \includegraphics[width=\linewidth]{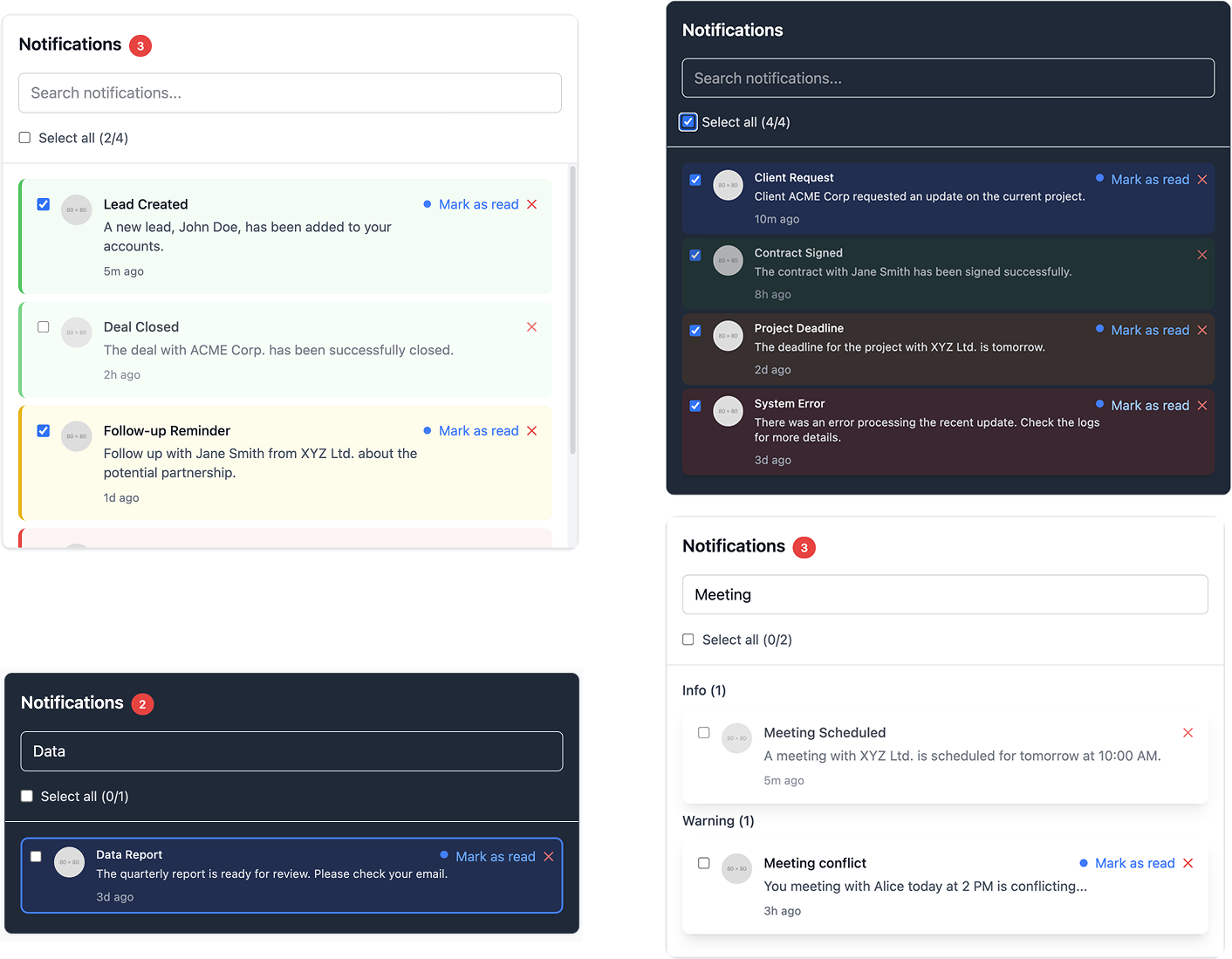}
    \caption{\textbf{Variations grounded on sampling instructions}. \textmd{\tool instantiates variations for a selectable and searchable notification center for a CRM app. \alex notices that all the variations contain notification text based on CRM-specific data, which is searchable and selectable.}}
    \Description{This image shows a screenshot of four different variations of the NotificationCenter component. \tool instantiates variations that are searchable, selectable, and data specific to the CRM domain.}
    \label{fig:forecast}
\end{figure}

\begin{figure}
    \centering
    \includegraphics[width=0.6\linewidth]{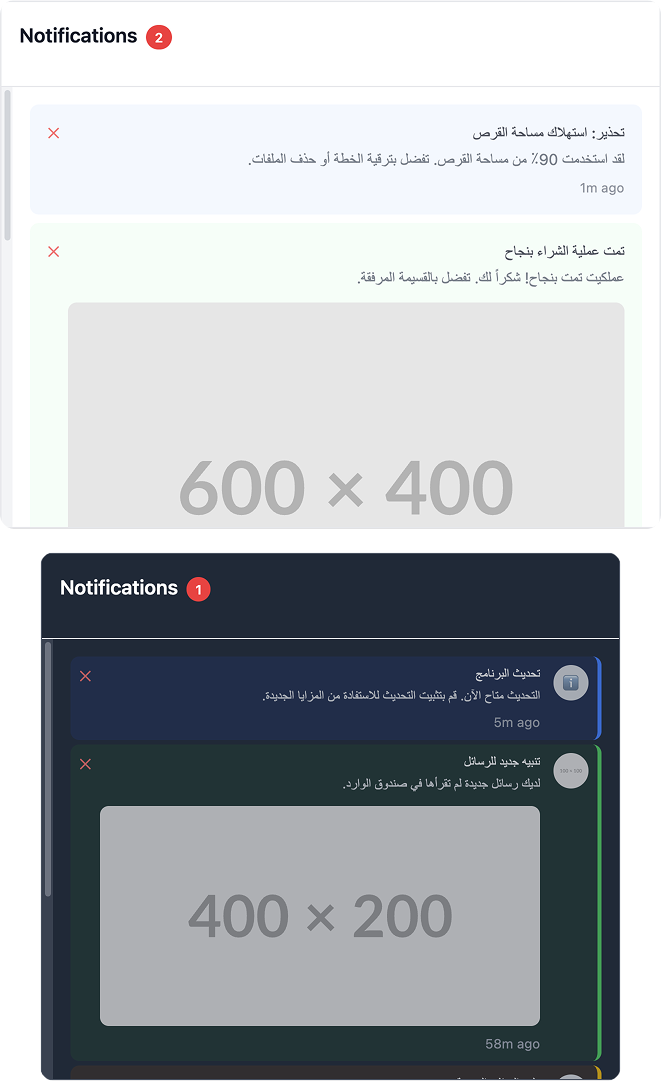}
    \caption{\textmd{\textbf{\alex uses \tool to instantiate more examples to answer two questions:} \say{Does it support right-to-left languages?} and \say{Does it support rendering media?}}}
    \Description{This image shows a screenshot of two more variations of NotificationCenter. \alex instantiates this by providing a custom sampling instruction to answer two questions: does it support right-to-left languages? and does it support media rendering?}
    \label{fig:compactdark}
\end{figure}


Moving from browsing to targeted exploration, \alex consults the \emph{coverage analysis} panel (\Cref{fig:teaser}~\protect\ccircle{C}) which highlights which properties have been exercised and which remain under-represented. The panel shows that sorting and grouping have been explored, but there are no examples with \smalltt{selectable} enabled. Still curious about searchable workflows, she uses the instruction box (\Cref{fig:teaser}~\protect\ccircle{B}) to request a focused slice of the design space: \say{Show me a searchable and selectable notifications for a CRM dashboard}. In seconds, \tool responds with targeted variations (\Cref{fig:forecast}) that combine \texttt{showSearch} and \smalltt{selectable}, populated with CRM-specific messages (e.g., client requests, ticket escalations, renewal reminders). Since the examples are both visually distinct and semantically grounded, \alex can quickly compare how property choices affect the layout; inspect individual variations and edit the code to understand how small property changes cascade from the container to the rendered notifications (\Cref{fig:inspect}).

Finally, \alex returns to the edge cases that previously blocked her adoption. She asks \tool to \say{Generate variations with Arabic notifications, and provide notifications with images}. The resulting examples reveal that the component supports RTL rendering, controlled via \texttt{enableRTL}, and demonstrate media layouts using placeholder images (\Cref{fig:compactdark}). She can quickly adjust image dimensions to observe how the layout responds and confirm that the component remains usable under the constraints of her scenario. With evidence across both common and critical edge cases, \alex is confident that \texttt{NotificationCenter} will meet her needs, and she has a clearer sense of which properties matter for her integration.


\section{Background and Related Work}

We draw from several threads of research to provide background concepts and related work that inform and motivate our work. We begin with a background on \sbook, the platform on which our tool, \tool, is built as an add-on, and explain why it is an appropriate setting. We then discuss the two main concepts: mimesis and variations. Finally, we connect our approach to research on sampling and model-finding in the domains of software testing and formal methods. 

\subsection{Background on \sbook}

Component-driven development (CDD) has become a dominant paradigm in front-end engineering, where user interfaces are assembled from modular, reusable components rather than monolithic pages~\cite{cdd, yew2020design, suarez2019design}. Rooted in component-based software engineering (CBSE), CDD emphasizes loose coupling, encapsulation, and separation of concerns: components are developed as independently testable units with well-defined interfaces, enabling teams to scale UI work across engineers and products. Frameworks such as React, Angular, and Vue.js embody this philosophy, while industry design systems like Google’s Material Design~\cite{materialdesign} and Microsoft’s Fluent Design System~\cite{microsoftFluentDesign} demonstrate how component libraries and standards can be shared across applications.

\sbook~\cite{storybook}, a widely used open-source tool, operationalizes CDD by providing a dedicated ``component workshop'' where developers author, visualize, and test components in isolation through stories---declarative examples that specify a component in particular states. In practice, developers add \sbook alongside an existing codebase and create stories as they build: a button might be captured in primary/secondary/disabled variants; a data table might include loading, empty, error, and large-data states; a form might include validation and submission outcomes. \sbook then renders these states in a browsable UI, letting developers jump directly to a scenario without running the full application, navigating to a screen, and reconstructing the right conditions. This makes it easy for downstream developers who use the UI components to view a variety of ``instantiated'' states without any setup, and also encourages the author to enumerate numerous examples to provide an exhaustive list of documented variations. 

Beyond being a convenient viewer, \sbook turns stories into shared artifacts that support collaboration and automation across the front-end pipeline, and often serve as living documentation for how components are intended to look and behave across contexts. Teams frequently use Storybook during design reviews (quickly comparing variations), regression testing (treating stories as stable baselines for visual or behavioral checks), and accessibility evaluation (running audits story-by-story). Prior work on GUI testing (e.g., Memon’s GUITAR~\cite{memon2013guitar}) similarly emphasizes systematic coverage of a UI’s state space; however, \sbook typically relies on manual enumeration of stories, leaving gaps in exploring the full design space of a component. This limitation motivates our work on AI-driven generation with \tool, which aims to expand coverage while preserving the practical, story-centered workflows that make Storybook useful.

\subsection{Mimesis and Mimetic Variation}
Aristotle's concept of \emph{mimesis}---often translated inadequately as \say{imitation} or \say{representation}---has served as a foundational idea shared by most authors, philosophers in the classical period~\cite{poetryfoundationMimesisimitation, sorbom2002classical, uchicagoMimesis}. While Plato saw mimesis as inferior, Aristotle, in his work \emph{Poesis}, viewed it as a natural human impulse and a positive force, enabling audiences to connect with and find meaning in art. In particular, Aristotle sees mimesis not as mere copying but as a natural, structured way for humans to learn and experience empathy and catharsis. Mimesis occupies a middle ground---neither purely historical, nor abstract---it represents what might occur, allowing audiences to recognize, empathize, and reflect. The concept of \emph{mimesis} has been used in design to anthropomorphize systems and robots~\cite{Bartneck2004, choi2008usage, Fink2012}, for affecting narratives in role-playing games~\cite{mimesiseffect2016}, and even pervasive computing~\cite{Rateau2014}. 
This philosophical lineage and the practical applications underpin our framing of mimetic variation in UI components: like mimetic art, effective UI variations should neither be arbitrary nor purely literal, but instead reflect plausible, coherent, and expressive use, not just exploring possibilities, but doing so in ways that also \say{feel} credible and meaningful in real-world contexts.

\subsection{Examples and Variations for Design Space Exploration}
Prior research in HCI has demonstrated the value of leveraging multiple examples and parallel variations to explore a design space. In creative work, generating or reviewing multiple alternatives in parallel can lead to more diverse ideas and higher-quality outcomes. For instance, \citet{dow2011} found that parallel prototyping yielded more divergent solutions than serial iteration. Examples are also central to inspiration-seeking: \citet{Herring2009} showed that examples are a cornerstone of creative design practice, used to trigger ideas, communicate concepts, and overcome blank slate challenges. Tools that curate and present example collections operationalize this insight: interactive example galleries let users browse and adapt design instances (even computer-generated), lowering the barrier to experimentation and helping uncover innovative directions~\cite{Lee2010, Ritchie2011, Huang2020}. However, when example sets grow into large numbers of near-duplicates, users must forage among highly similar variants; \citet{ragavan2016foraging} studied how novice programmers navigate and evaluate hundreds of similar code variants and proposed refinements to Information Foraging Theory for such variant-rich settings. Complementing this account of variant foraging, \citet{kery2017variolite} introduced Variolite to support exploratory data-science programming via lightweight, local versioning of arbitrary code regions and rapid switching/comparison among alternatives.
Beyond supporting exploration and information foraging, examples and their variations also facilitate conceptual understanding. Variation Theory~\cite{marton2013learning, marton2014necessary} explains learning as discerning critical features of a concept by experiencing structured differences---varying one aspect while holding others constant. This principle has informed computing education, where varying teaching strategies or contrasting examples help students grasp difficult concepts~\cite{suhonen2008applications}. In HCI, systems like OverCode~\cite{Glassman2015overcode} visualize clusters of student programming solutions or API examples~\cite{classman2018examplore} to highlight conceptual variation and support pattern recognition.
Our work builds on this recognition of examples as drivers of divergence and inspiration, but differs by addressing the specific challenge of UI component exploration: instead of browsing a static gallery or retrieving curated artifacts, our system automatically generates distinguishing variations of UI components tailored to the developer's context. By embedding this capability into \sbook, an industry-standard tool, we extend prior work on example galleries into a workflow where examples are not merely browsed but actively sampled and instantiated within the design space of interactive components.

\subsection{UI Variations as Design Space Sampling}

Software testing has long framed test generation as a sampling problem over vast input spaces. Early work, such as McKeeman's differential testing of C compilers~\cite{mckeeman1998differential}, generated random inputs and compared outputs across implementations. Since exhaustive testing is infeasible, researchers have developed strategies to select representative inputs: random testing~\cite{pacheco2007feedback}, partitioning techniques that divide domains into equivalence classes and boundaries~\cite{gutjahr2002partition}, and combinatorial testing that ensures coverage of all $t$-way parameter interactions, as most failures stem from only a few interacting factors~\cite{kuhn2004software}. More recent approaches incorporate guidance, such as property-based testing~\cite{fink1997property, maciver2019hypothesis, paraskevopoulou2015foundational, goldstein2024property} or feedback-directed generation~\cite{pacheco2007feedback}, to better explore hard-to-reach cases. These methods have also been applied to GUIs, where event sequences are sampled systematically to uncover interaction faults~\cite{memon2003gui, yuan2010gui}.
We adopt this sampling lens for generating UI component variations: each component exposes properties that span a large design space of layouts, styles, and content. Like test case generation, our approach samples \emph{interesting} combinations of property values to traverse this space. Unlike tests, however, variations are directly presented to users, requiring aggressive pruning to avoid overload while still providing a \emph{birds-eye view} of the design space. This also introduces an additional constraint: variations must be \emph{mimetic}---combinations that are plausible, natural, and contextually relevant.

\subsection{Example Generation with Model Finding}

In formal methods, model finders use SAT or SMT solvers to generate concrete instances that satisfy a specification~\cite{daniel2002alloy, mccune2001mace, Leuschel2003, Blanchette2010, Dennis2006, Dolby2007}. Rather than abstract proofs, they return tangible examples or counterexamples that help developers validate and explore the design space~\cite{Danas2017}. This process is analogous to exploring variations in UI components: in both cases, the goal is to concretize an abstract specification into representative, interpretable instances that expose the possibilities and limitations of the design. Just as tools like Alloy help explore bounded design spaces through automatic instance generation~\cite{daniel2002alloy}, variation generators for UI can reveal the span of potential instantiations within component libraries by systematically sampling from parameterized spaces.
While the core benefits of model finders are promising in theory, users have often found these to be less helpful in practice~\cite{zave2015practical, mansoor2023empirical, Danas2017}. A long-held belief is that there are too many solutions, which can overwhelm the user. As a result, there is a whole body of work dedicated to trying to reduce the number of solutions~\cite{Sullivan2021, Sullivan2017, Porncharoenwase2018}, often by creating additional constraints on the solution. We face a similar problem with UI variation sampling: most of the variations often contain unrealistic inputs, or they are minor tweaks of existing variations. In this paper, by using LLMs with clever prompts and context engineering, we make the sampling constrained to having \emph{naturalistic} and \emph{domain-relevant} values for component properties.


\section{Formative Study} \label{sec:formative}

We conducted a formative study with $9$ front-end developers at a large technology company to understand the process of authoring and maintaining UI component libraries using tools like \sbook, and the challenges that accompany it. Participants were recruited via an internal mailing list. Participants had an average of $9.2$ years ($min = 3.5$, $max = 16$, $\sigma = 4.11$) of front-end engineering experience; and an average of $4.3$ years using tools like \sbook ($min = 1$, $max = 9$, $\sigma = 2.3$). We conducted semi-structured interviews lasting approximately $30$ minutes, allowing participants to describe their workflows, tooling practices, and pain points. Sessions were recorded, transcribed, and thematically analyzed.

\subsubsection{Challenges}

Participants described \sbook as a shared resource that supports day-to-day component development and is routinely consulted as reference material for integrating components into a variety of products and workflows. Eight participants (except P6) noted that their \sbook instances were accessed by external teams, and two participants  (P2, P5) said their library was used outside their organization. 
Across these settings, participants emphasized that the value of \sbook workbench is largely determined by what it provides to the consumers: reliable, discoverable examples that help them understand how to adopt components without needing deep familiarity with the underlying codebase.

\paragraph{Variant coverage is essential but under-specified.}
Participants emphasized that story variations were critical for downstream understanding, since \say{these are really complicated components with a long list of properties and options. And without reading the code, users [would not] be aware of them.} (P1). Variants were especially important for developers unfamiliar with the codebase of the components they are using, since \say{it's just hard to get to that variant of the component inside the real app} (P2). Yet, participants described variant selection as informal and inconsistent, often driven by \say{the subjective choice of the [\sbook author] to actually decide on stories for a component} (P9). Participants noted that these choices frequently did not reflect the use cases of the developers consuming the component library. As a result, \sbook stories could only communicate a narrow slice of the component's behavior, leaving the developers uncertain about its use for their specific scenario. 
We asked participants to share the \sbook workbenches they used and gained access to five enterprise component workbenches containing a total of 253 components. We inspected all five workbenches and found that, on average, each component had 3.11 variations (stories) specified ($\sigma = 2.34$).

\paragraph{Authoring variants is time-intensive and competes with feature work.}
Six participants who also authored \sbook workbenches in the past said that constructing useful story variants was among the hardest parts of maintaining a component library. One participant described routinely limiting themselves to a small, representative set: \say{I try to target like three to five stories generally, and it always takes a lot of time, maybe 15-20 \% of the total time} (P4). Participants characterized this work as high-effort because it requires both implementation labor (setting up properties and states) and design judgment (deciding what is worth documenting). 
Yet this documentation work was commonly deprioritized relative to product delivery: \say{documentation is probably not on the top of the priorities when it comes to feature development, and product support and documentation take the heat. It's always the last you work on} (P9). This made the story coverage fragile: even when good variants were initially authored, the development pace often made it difficult to keep StoryBooks synchronized.

\paragraph{Complex, data-specific components are hard to instantiate for every use case.}
Four participants highlighted that instantiating realistic variants is particularly challenging for components that depend on custom backend data. As one participant put it, \say{for complex components, it is very complex to set up mock data. We use Redux to fetch data, but it is annoying to set up --- this limits the complexity of the component inside \sbook} (P2). Participants mentioned that mock data is not merely a convenience for authors: it is often essential for developers to understand how a component is intended to be used, what inputs it expects, and what realistic states and behaviors look like in practice. At the same time, because \sbook often serves a wide variety of developers with divergent products and workflows, it is impossible to represent every use-case in the stories. This often leaves consumers of component libraries relying on incomplete and abstract examples.

\subsubsection{Opportunities}

Participants suggested several opportunities to improve the status quo for \sbook consumers and authors. First, they emphasized the value of automating variation generation: \say{automating this [story generation] step would be very helpful}, particularly if it could generate \say{the best examples to showcase} (P3). Second, participants wanted to explore beyond a fixed set of stories, for example, by specifying an integration requirement and asking \sbook to surface a matching variant: \say{if you can just prompt to show a variation of a component that’s based on this requirement, that would be super helpful} (P7). Third, participants wanted support to \say{be able to test with mock data given the API or parameters} (P7), especially for data-heavy components where credible mock data is time-consuming to author (e.g., \say{components like select or table, you might have a list of 50 states, or an address. There are use cases where having some good dummy data would be very helpful} (P8)). Finally, P4 wanted \say{to be able to edit things [\sbook] has generated, so I can tune what I need and copy-paste to use it} (P4).

Many participants noted they already use generative AI tools for software engineering. However, five participants explicitly said that the current generative AI tools are not effective in instantiating good UI variations: \say{It is a step back to use [generative AI] for story generation. It doesn't understand the component tree. I will have to re-prompt a couple of times, and I'm still manually going and editing it further} (P8). They were also skeptical of using these tools to instantiate story variants: \say{It really depends on the business logic. Not all combinations make sense. It'll be helpful if it can understand the properties and generate useful combinations} (P2).


\section{The \tool System}

We developed \tool, implemented as a \sbook add-on, to support exploration of distinguishing variations of UI components. We first describe the system's design motivations, informed by our formative study, and then describe the framework behind \tool for dynamically instantiating component variations.

\begin{figure*}
    \centering
    \includegraphics[width=1\linewidth]{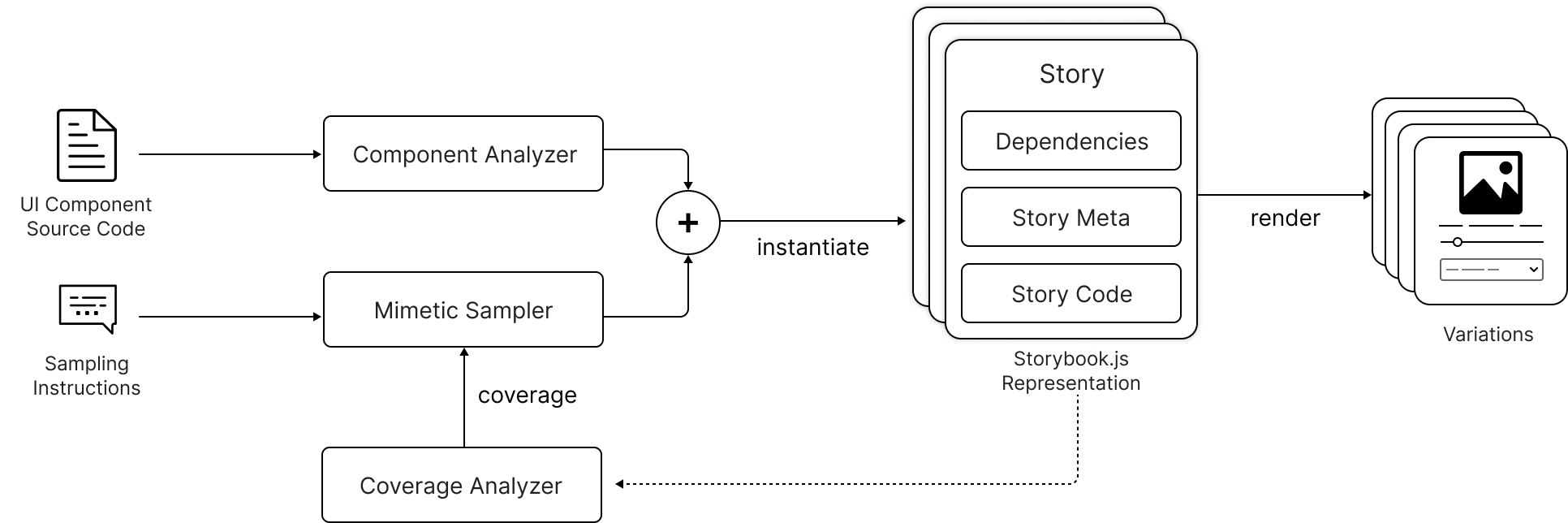}
    \caption{\tool system architecture. \textmd{The \componentAnalyzer{} uses symbolic techniques to retrieve a list of highly visually-impactful properties of the UI component. The \coverageAnalyzer{} measures the coverage of the generated variations to inform the sampler. Finally, the \naturalSampler{}, powered by an LLM, uses the component analysis and the coverage metrics along with an optional sampling instruction from the developer to generate distinguishing variations.}}
    \Description{This image shows a screenshot of \tool system architecture. The \componentAnalyzer{} uses symbolic techniques to retrieve a list of highly visually-impactful properties of the UI component. The \coverageAnalyzer{} measures the coverage of the generated variations to inform the sampler. Finally, the \naturalSampler{}, powered by an LLM, uses the component analysis and the coverage metrics along with an optional sampling instruction from the user to generate distinguishing variations.}
    \label{fig:system}
\end{figure*}

\subsection{Design Motivations}

Our formative study revealed recurring challenges in authoring and exploring story variations in UI component libraries. From our observations, we distilled two design motivations:

\begin{description}
    \item[DM1:] \textbf{Promote exploration of variations to a first-class activity.} The main purpose of documentation workbenches like \sbook is to let developers explore components to understand how and where they can be used in practice. Yet participants emphasized that creating story variants was manual, ad hoc, and \say{one of the hardest tasks} of maintaining a component library (P4, P9). As a result, the limited hand-authored stories often restricted the discovery of the full design space. Any solution should therefore treat exploration as a first-class activity, providing dedicated affordances that make it easy to surface and compare a broad range of variants. 
    At the same time, participants noted that not all property combinations were meaningful in practice, and very much depend on business logic (P2). Because shared component libraries are isolated from application-specific logic, developers wanted a way to guide exploration toward their particular use cases and requirements (P4, P7). Thus, exploration must be open-ended yet steerable, balancing broad discovery with specific goals.

    \item[DM2:] \textbf{Instantiate relatable components in domain-specific scenarios.} 
    Another challenge in creating meaningful story variations is populating components with data that feels \emph{relevant} to the intended use (P2, P4, P7, P8). Placeholder data that is too generic can make a component appear generic or even misleading, while data that reflects domain semantics helps developers better judge how the component will behave in practice for their use case. Since shared libraries often serve diverse application domains, developers are not always able to relate to the pre-authored component instantiations, and the engineering effort to set up data mocks is laborious. Therefore, any generated story variation should contain relevant, domain-faithful data that makes the examples feel plausible and relatable to the developer's context.
\end{description}

\subsection{System Architecture}

\tool\ consists of three primary modules: \componentAnalyzer{}, \naturalSampler{}, and \coverageAnalyzer{} (\Cref{fig:system}). These modules leverage symbolic program analysis (\componentAnalyzer{} and \coverageAnalyzer{}) and large language models (\naturalSampler{}) to generate distinguishing UI component variations.

\subsubsection{ProductCard example}

To first ground the discussion in this section with a concrete running example, we introduce the \smalltt{ProductCard} component shown in \Cref{lst:productcard}, through which we motivate the design rationale and operation of each module. \smalltt{ProductCard} represents a typical component used for depicting a purchasable product in an online storefront. It exposes typical properties such as \smalltt{title}, \smalltt{price}, and product \smalltt{image}, along with specialized properties that control visual styling such as layout density (\smalltt{variant}) and color (\smalltt{theme}).

\begin{lstlisting}[
  language=TSX,
  style=code,
  caption={The code for \smalltt{ProductCard} component used as the running example.},
  label={lst:productcard}
]
type ProductCardProps = {
  variant?: "summary" | "detailed"; // layout structure
  title: string;                    // main content
  price: number;                    // numeric content
  imageUrl?: string;                // conditional rendering
  theme?: "light" | "dark";         // styling impact
  showBadge?: boolean;              // conditional rendering
  borderStyle?: "solid" | "dashed"; // low visual impact
};

export const ProductCard: React.FC<ProductCardProps> = ({
  variant = "summary", title, price, imageUrl, 
  theme = "light", showBadge = false, borderStyle = "solid",
}) => {
  return (
    <div className={`product-card ${theme} 
        border-${borderStyle}`}>
      {variant === "detailed" && imageUrl && 
        (<img src={imageUrl} alt={title} />)}
      <h2>{title}</h2>
      {variant === "detailed" && 
        <p className="price">${price}</p>}
      {showBadge && <span className="badge">New</span>}
    </div>
  );
};
\end{lstlisting}

\subsubsection{ComponentAnalyzer module}
\label{sec:component-analyzer-module}

%
The \componentAnalyzer{} estimates the visual impact of different component properties on the overall appearance of the rendered component UI. We define our notion of visual impact more precisely below when we discuss our visual impact scoring function;
intuitively, \componentAnalyzer{} ranks the impact of property changes from coarse to fine, and considers coarse-grained changes (e.g., modification of component layout or structure) to be greater than fine-grained changes (e.g., modification of detailed aesthetics such as individual CSS values). 

\componentAnalyzer{} operates over abstract syntax trees (ASTs) as an intermediate representation, so its first stage parses the component's implementation in the form of React and TypeScript source code.
A second stage performs a static information flow analysis over the AST to trace flow from property values to syntactic elements that directly influence the visual appearance:
\begin{enumerate*}
  \item TSX or JSX expressions,
  \item conditional rendering blocks,
  \item CSS property value assignments, and
  \item inline style assignments.
\end{enumerate*}
A final stage ranks the syntactic elements thus discovered, using the following classifications (in descending order of visual impact):

\begin{description}
\item[Structure:] Expressions that determine which DOM subtrees should be rendered. This includes expressions that implement subcomponent rendering, conditional rendering, or dynamic subcomponent selection. For example, in \smalltt{ProductCard} (\Cref{lst:productcard}), the \smalltt{variant}, \smalltt{showBadge}, and \smalltt{imageUrl} properties all flow to conditionally-rendered TSX expressions.
\item[Content:] Expressions that interpolate property values into rendered content. This includes expressions that implement list/map rendering for array properties or dynamic text content. For example, the \smalltt{title} and \smalltt{price} properties of \smalltt{ProductCard} flow to the inner text content for the \smalltt{<h2>} and \smalltt{<p>} tags, respectively.
\item[Styling:] Expressions that determine CSS styling, such as class names, attribute values, and inline styles. For example, the \smalltt{theme} and \smalltt{borderStyle} properties of \smalltt{ProductCard} both interpolate into \smalltt{className} attribute assignments.
\end{description}

\noindent
We call the expressions classified above \emph{visually impactful code contexts} (\emph{vi-contexts} for brevity) and denote by $V(p)$ the set of vi-contexts found for property $p$.

To compute a numerical score summarizing the visual impact for each property $p$, we then combine two factors:
\begin{enumerate*}
\item the type of each vi-context $v \in V(p)$ (from which we compute a \emph{base impact factor}), and
\item the total number of vi-contexts $|V(p)|$ (from which we compute an \emph{impact coefficient} that potentially boosts the score).
\end{enumerate*}

\paragraph{Base impact factor}
The base impact factor $B(p)$ corresponds to the highest-ranked vi-context discovered for $p$ during flow analysis (recall that vi-contexts are ranked from coarse- to fine-grained)
\[B(p)=\max_{v \in V(p)}b(v)\]
where $b(v)$ associates each vi-context $v$ with a positive score. We chose a uniform heuristic that starts from 100 (maximum base impact) and reduces by 20 for each lower-ranked vi-context (i.e., $100$, $80$, and $60$ for structure, content, and styling vi-contexts, respectively).

\paragraph{Impact coefficient}
Since a property may flow to multiple vi-contexts (e.g., $|V(p)| > 1$), we additionally apply an upward adjustment called the impact coefficient $C(p)$ based on the number of times a property appears in a vi-context. Intuitively, we rely on the heuristic that properties that affect many parts of the component are more impactful than those that only affect one:
\[C(p)=1 + (1 - \exp(-n/10))\]
The exponential decay function rewards frequently used properties while diminishing the contribution of each additional usage to prevent very frequently-appearing properties from dominating all others.

\paragraph{Visual impact scoring function.}

Putting it together, the visual impact scoring function $I(p)$ is simply the product of the base impact factor and the impact coefficient.
\[I(p) = C(p) \times B(p)\]
Note that since $C(p) \geq 1$, $I(p) \geq B(p)$. In other words, the final visual impact score can never be reduced below the base impact factor.

Finally, to determine the set of visually-impactful properties, we apply a simple cutoff and consider only those properties with $I(p) \geq 100$, while the rest are filtered out as low-impact properties.

\subsubsection{MimeticSampler module}
\label{sec:mimetic-sampler-module}  

While \componentAnalyzer{} identifies the subset of properties that are most visually impactful, \naturalSampler{} generates concrete property values that lead to mimetic and distinguishing variations of the component as a whole.
At its core, \naturalSampler{} employs a large language model (LLM) for \emph{mimetic sampling}: generation of candidate property values that follow human conventions and domain semantics.

\paragraph{Prompt construction}
The input prompt contains a structured representation of the component interface schema that includes
\begin{enumerate*}
  \item property names,
  \item inferred property types,
  \item visual impact scores as computed by \componentAnalyzer{}, and
  \item optionally, extracted source code snippets for vi-contexts corresponding to the highest base impact scores.
\end{enumerate*}
We additionally include instructions that guide the LLM towards three objectives: 
\begin{enumerate}
  \item \emph{Mimesis}: ensure that generated values are relevant to the domain and aligned with real-world conventions [\textbf{DM2}], 
  \item \emph{Distinctness}: guide generation toward under-sampled parts of the design space via feedback from the \coverageAnalyzer{} (\Cref{sec:coverage-analyzer-module}) [\textbf{DM1}], and
  \item \emph{User Instruction-Following}: follow the user's special sampling instruction, if any, with high priority [\textbf{DM1}].
\end{enumerate}
We have also conditioned the prompt to prioritize the properties with high visual impact (Appendix~\ref{ap:prompt}).

Finally, \naturalSampler{} supplies this prompt along with a query to generate all property values jointly in a single response (we do not perform individual requests for each property). Intuitively, we do this because property values are frequently correlated, so the model should generate \emph{sets of values} that are mutually compatible (e.g., \smalltt{imageUrl} should only be assigned when \smalltt{variant="detailed"}, as the \smalltt{"summary"} variant does not render an image).

\subsubsection{CoverageAnalyzer module}
\label{sec:coverage-analyzer-module}

The \coverageAnalyzer{} measures how well generated stories explore a component's design space. It reuses the machinery of the \componentAnalyzer{} to extract property values from story code and compare them against the set of visually impactful properties.  

\paragraph{Measuring coverage.}  
Coverage is defined with respect to four property types:
\begin{itemize}
    \item \emph{Categorical properties:} Categorical properties may take one of several values in an enumeration. Coverage measures the fraction of possible values actually observed across generated stories. For example, \smalltt{variant} may take on either \smalltt{"summary"} or \smalltt{"detailed"} as values.
    \item \emph{Boolean properties:} Boolean properties may be \smalltt{true} or \smalltt{false}. Full coverage requires both \smalltt{true} and \smalltt{false} values to appear, either explicitly or implicitly via a default argument. For example, \smalltt{showBadge} may take on either \smalltt{true} or \smalltt{false} values to enable or disable a badge icon.
    \item \emph{String or numeric properties:} Since strings and numerical values have unbounded domains, we use heuristics to estimate coverage. In the case of string properties, full coverage requires that a property contains at least three unique values, with at least one long value ($>50$ characters). For numerical properties, full coverage requires that a property take on at least three unique values.
    \item \emph{Object properties:} For nested objects containing deep property configurations, we recursively derive coverage using the measures just described.
\end{itemize}

Formally, for a property $p$ with domain $D(p)$ and observed set $O(p)$ across generated stories, coverage is defined as:  
\begin{equation*}
C(p) = \frac{|f(O(p))|}{|f(D(p))|} 
\end{equation*}
where $f(\cdot)$ maps values into semantic equivalence classes (that is, string categories or numeric buckets). A property is considered fully covered only when $C(p) = 1$.

\paragraph{Closing the loop.}  
As mentioned in \Cref{sec:mimetic-sampler-module}, the \coverageAnalyzer{} feeds its results back to the \naturalSampler{}. Specifically, the \naturalSampler{} constructs prompt instructions from these results to request generation of under-explored property values, prioritizing properties with high visual impact, to close coverage gaps (e.g., the absence of \smalltt{variant="detailed"} stories).
%

\subsection{\tool UI}

The user interface of \tool is organized to support broad exploration while still allowing users to drill into details when needed, using a consistent spatial layout: a central canvas for rendering all component instantiations and a right-hand panel for structured information and controls. 

\subsubsection{Visual Overview} 
Interaction begins in the \emph{Visual Overview}, which renders all generated variants for a selected component on a single canvas (\Cref{fig:teaser}~\protect\ccircle{A}). This view enables rapid scanning, side-by-side comparison, and iterative expansion of the example set. 
Users can also provide optional \emph{sampling instructions} through an instruction textbox (\Cref{fig:teaser}~\protect\ccircle{B}), steering the sampler toward specific areas of interest [\textbf{DM1}]. 
The side panel provides a \emph{Coverage Analysis} that enumerates property values represented across the current examples and highlights values not yet explored (\Cref{fig:teaser}~\protect\ccircle{C}). When users request additional examples, \tool uses this summary to guide generation toward uncovered regions of the design space, keeping exploration systematic rather than ad hoc. The coverage view dynamically updates as more variants are generated and added to the canvas. 

\begin{figure*}
    \centering
    \includegraphics[width=1\linewidth]{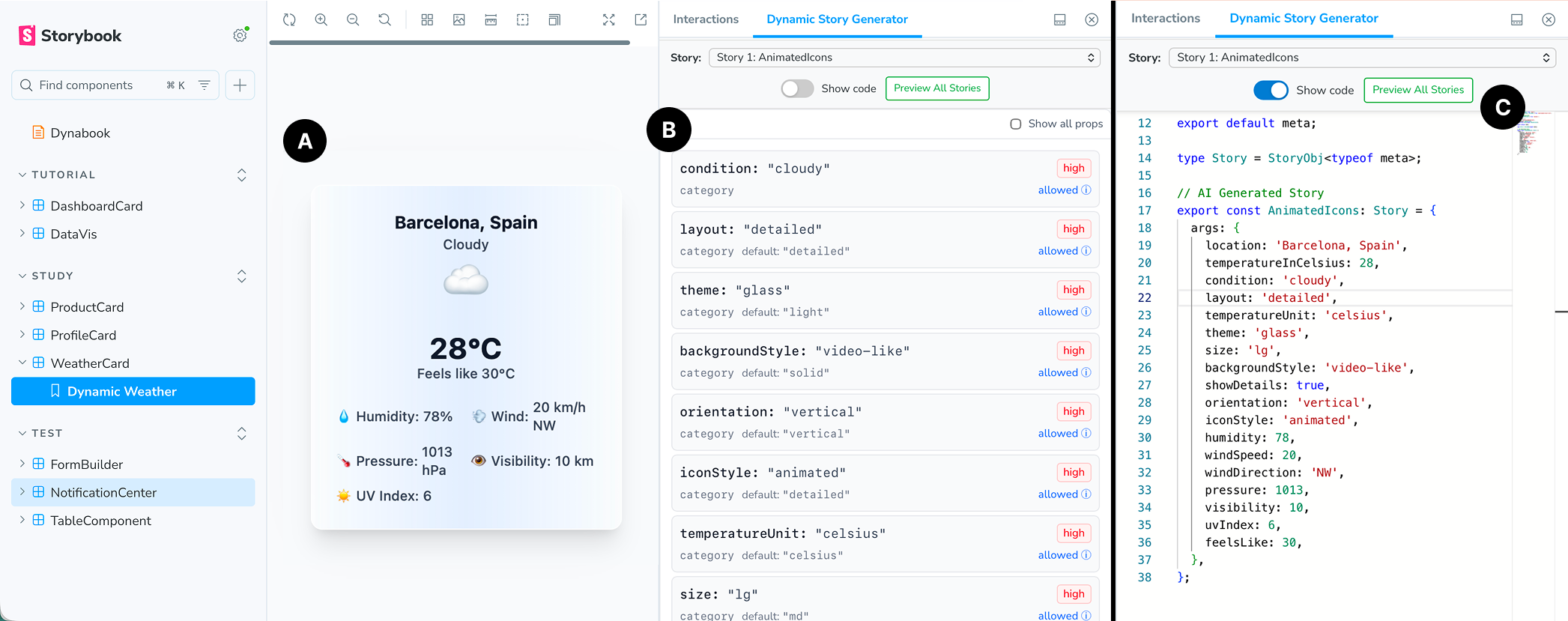}
    \caption{\tool's inspect UI. \textmd{The developer can choose to inspect a single variant in detail using \tool's inspect feature. \protect\ccircle{A} The center pane renders the selected variant. \protect\ccircle{B} The right pane shows the property values that instantiate the selected variant. \protect\ccircle{C} The user can also choose to inspect the \sbook code for the variant, and optionally edit the properties to immediately see the changes rendered in the center pane.}}
    \Description{This image shows a screenshot of \tool's inspect UI. The user can choose to inspect a single variant in detail using \tool's inspect feature. The center pane renders the selected variant. The right pane shows the property values that instantiate the selected variant. The user can also choose to inspect the \sbook code for the variant, and optionally edit the properties to immediately see the changes rendered in the center pane.}
    \label{fig:inspect}
\end{figure*}

\subsubsection{Inspect}
Selecting any variant in the overview brings it into focus for closer inspection. The chosen example is rendered in the center (\Cref{fig:inspect}~\protect\ccircle{A}), while the side panel lists its instantiated property values (\Cref{fig:inspect}~\protect\ccircle{B}). A \emph{Show Code} toggle reveals the underlying \sbook snippet in an inline editor, allowing users to connect the rendered output with its specification (\Cref{fig:inspect}~\protect\ccircle{C}). Property values can be modified directly in the editor with immediate visual feedback, offering a lightweight way to test how changes affect a component's appearance and behavior.

By moving back and forth between visual overview and inspection, users can quickly browse and compare many generated variants, while still being able to focus on any single example to understand how specific property values and code lead to its rendering. Transitions between overview and detail are lightweight and remain within the same page.

\subsection{Implementation}

We implemented \tool\ as a \sbook addon using React and TypeScript, bundled with Vite. The system integrates large language models for sampling by querying the OpenAI API~\cite{openaiPlatform}. Our current implementation uses \smalltt{GPT-4o}~\cite{openaiHelloGPT4o}, which offers a practical balance between accuracy and speed. The sampling approach is model-agnostic: other state-of-the-art models (e.g., \smalltt{GPT-5}~\cite{openaiIntroducingGPT5} or \smalltt{Claude 4}~\cite{anthropicIntroducingClaude}) could be used as long as the API supports strict JSON output, which is necessary to keep sampled values type-accurate.  
For rendering, we use the \smalltt{react-html-parser} library~\cite{npmjsHtmlreactparser} to dynamically attach and detach HTML elements at runtime. The sampled property configurations are then composed into new stories using \sbook's internal APIs. Each generated variation is automatically wrapped as a React component, making it directly explorable within the design environment. 

The \texttt{ComponentAnalyzer} and \texttt{CoverageAnalyzer} are implemented using the Babel packages that provide off-the-shelf parsing for TypeScript. The \smalltt{@babel/parser} package is used to parse the TSX/JSX code into an abstract syntax tree (AST) by both the analyzers. \texttt{ComponentAnalyzer} also uses the \smalltt{@babel/traverse} package to implement a custom traversal algorithm over the parsed AST to compute the visual impact scores, and the \smalltt{@babel/types} package to identify variable types. The \texttt{ComponentAnalyzer} does two passes over the AST to compute the visual impact scores. The first discovery pass extracts component properties from function parameters, identifies state variables and function properties, collects TypeScript interfaces and type definitions, handles renamed props, and detects whether the component uses destructured properties. The final stage is when the analyzer assigns visual impact scores based on the observed usage patterns.


\section{\tool Evaluation}

We conducted both a technical evaluation and a user study of \tool. The technical evaluation assesses the accuracy of \componentAnalyzer{} in classifying the visual impact of component properties. The user study examines how distinguishing variations support developers in exploring component design spaces and understanding component behavior.

\subsection{Technical Evaluation}
To assess the effectiveness of \componentAnalyzer{} at estimating the visual impact of component properties, we conducted a technical evaluation on ten complex components from Ant Design---a popular open-source React UI component library~\cite{ant-design}. Our goal was to quantify how accurately \componentAnalyzer{} assigns each property to one of three visual-impact levels (High/Medium/Low) when compared against human labels.

\subsubsection{Setup}
We selected ten complex Ant Design components with the highest number of visual properties: \smalltt{alert}, \smalltt{breadcrumb}, \smalltt{dropdown}, \smalltt{form}, \smalltt{menu}, \smalltt{modal}, \smalltt{progress}, \smalltt{steps}, \smalltt{time-picker}, and \smalltt{timeline}. For each component, we enumerated its configuration properties from the source code and classified its visual impact. In total, the evaluation set includes 172 property instances across the ten components (Table~\ref{tab:techeval}).
We defined visual impact as the expected magnitude of \emph{visible} change produced by varying a property value: (1) \textit{High} substantially changes appearance, layout, or dominant visual structure; (2) \textit{Medium} produces noticeable but localized visual differences (e.g., secondary styling changes); and (3) \textit{Low} results in subtle or minor visible changes. Two of the authors labeled each property using this rubric (Cohen's $\kappa=0.8$).
We ran \componentAnalyzer{} on the same set of properties and compared predicted labels to ground truth. We report \textit{micro-averaged accuracy}---the fraction of correctly classified property instances across the full dataset---as well as per-component accuracy (Table~\ref{tab:techeval}).

\begin{table*}[h]
\centering
\small
\caption{Accuracy of \smalltt{ComponentAnalyzer} in classifying visual impact (High/Medium/Low) of properties across ten Ant Design React components. \textmd{Our method achieves 83.1\% accuracy (143/172), with performance generally stable across components, suggesting the classifier transfers reasonably well across diverse UI patterns.}}
\label{tab:techeval}
\begin{tabular}{l|llll|l|l}
\multirow{2}{*}{\textbf{Component}} & \multicolumn{4}{l|}{\textbf{\begin{tabular}[c]{@{}l@{}}Number of properties \\ (visual impact)\end{tabular}}}        & \multirow{2}{*}{\textbf{\begin{tabular}[c]{@{}l@{}}Classification Accuracy\\ (ComponentAnalyzer)\end{tabular}}} & \multirow{2}{*}{\textbf{Matches (n/N)}} \\ \cline{2-5}
                                    & \multicolumn{1}{l|}{\textbf{High}} & \multicolumn{1}{l|}{\textbf{Medium}} & \multicolumn{1}{l|}{\textbf{Low}} & \textbf{Total} &                                                                                                                 &                                                                     \\ \hline
alert                               & \multicolumn{1}{l|}{9}             & \multicolumn{1}{l|}{3}               & \multicolumn{1}{l|}{8}            & 20             & 80.0                                                                                                            & 16/20                                                               \\
breadcrumb                          & \multicolumn{1}{l|}{4}             & \multicolumn{1}{l|}{0}               & \multicolumn{1}{l|}{7}            & 11             & 81.8                                                                                                            & 9/11                                                                \\
dropdown                            & \multicolumn{1}{l|}{7}             & \multicolumn{1}{l|}{3}               & \multicolumn{1}{l|}{13}           & 23             & 87.0                                                                                                            & 20/23                                                               \\
form                                & \multicolumn{1}{l|}{6}             & \multicolumn{1}{l|}{7}               & \multicolumn{1}{l|}{8}            & 21             & 76.2                                                                                                            & 16/21                                                               \\
menu                                & \multicolumn{1}{l|}{5}             & \multicolumn{1}{l|}{5}               & \multicolumn{1}{l|}{4}            & 14             & 78.6                                                                                                            & 11/14                                                               \\
modal                               & \multicolumn{1}{l|}{6}             & \multicolumn{1}{l|}{8}               & \multicolumn{1}{l|}{11}           & 25             & 88.0                                                                                                            & 22/25                                                               \\
progress                            & \multicolumn{1}{l|}{8}             & \multicolumn{1}{l|}{4}               & \multicolumn{1}{l|}{3}            & 15             & 80.0                                                                                                            & 12/15                                                               \\
steps                               & \multicolumn{1}{l|}{14}            & \multicolumn{1}{l|}{2}               & \multicolumn{1}{l|}{5}            & 21             & 85.7                                                                                                            & 18/21                                                               \\
time-picker                         & \multicolumn{1}{l|}{3}             & \multicolumn{1}{l|}{2}               & \multicolumn{1}{l|}{3}            & 8              & 75.0                                                                                                            & 6/8                                                                 \\
timeline                            & \multicolumn{1}{l|}{8}             & \multicolumn{1}{l|}{0}               & \multicolumn{1}{l|}{6}            & 14             & 92.9                                                                                                            & 13/14                                                              
\end{tabular}
\end{table*}

\subsubsection{Results}
Overall, \componentAnalyzer{} correctly classified 143 out of 172 properties, yielding an accuracy of \textbf{83.1\%} (Table~\ref{tab:techeval}). Performance was generally consistent across components, ranging from \textbf{75.0\%} (\smalltt{time-picker}) to \textbf{92.9\%} (\smalltt{timeline}), suggesting the approach generalizes across diverse component types rather than benefiting from a single favorable case.

Since \componentAnalyzer{} is intended to support prioritization when generating variations, we also inspected \emph{extreme} misclassifications (High$\rightarrow$Low or Low$\rightarrow$High). These occurred in only 3 properties total (1.7\%), with zero extreme mismatches in eight components. This pattern suggests that when \componentAnalyzer{} errs, it typically confuses adjacent categories rather than reversing impact. However, edge cases remain for properties whose visual effects are highly contextual or primarily semantic.

\subsection{User Evaluation}
The main aim of the user study is to answer the following research questions:
\begin{description}
    \item[RQ1.] How do distinguishing variations support participants in exploring the design space of UI components? 
    \item[RQ2.] How do participants understand the properties and component behavior as they explore?
    \item[RQ3.] How do participants perceive mimetic variations and their usefulness in practice?
    \item[RQ4.] What challenges or limitations do participants encounter when engaging with instantiated variations?
\end{description}

\subsubsection{Participants}
We recruited $12$ professional front-end developers ($P1-P12$) by posting study invitations to internal communications channels at a large technology company. Participants self-reported an average of $8.4$ ($min=1$, $max=25$, $\sigma=6.74$) years of experience as web front-end developers. Eleven participants reported that they have used generative AI tools for work, although this was not a specific requirement to participate in the study. Four of the 12 participants had also participated in the formative study. As part of the recruitment process, participants were also informed they would receive \$25 meal vouchers for their participation. The total study time was 60 minutes.

\subsubsection{Procedure}

The user study has four stages: onboarding, exploration tasks, a post-task questionnaire, and an informal interview.

\paragraph{Onboarding and tutorial (15 minutes)} Participants were introduced to the study goals and the \tool interface. After informed consent, to familiarize participants with the interaction paradigm, we began with a guided tutorial using a simple UI component. The tutorial demonstrated both the visual overview and the inspection features of the tool with a separate UI component specifically assigned for the tutorial.
    
\paragraph{Open-ended exploration tasks (30 minutes)} After the tutorial, participants completed two open-ended exploration tasks using \tool, during which they were asked to think aloud. In each task, participants were instructed to generate variations for a selected UI component implemented in React (see~\Cref{ssec:studycomp} for details). They were encouraged to create as many variations as they wished until they felt confident that they had covered all possibilities of the component. After each iteration of generating examples, participants rated the variations on three metrics: (1) how natural and realistic they were, (2) how distinct they were from one another, and (3) how well they covered the component's possibilities (including earlier examples). Here we use \emph{natural} and \emph{realistic} as a proxy for \emph{mimetic}, since participants may not be familiar with the term. Ratings were provided on a 5-point Likert scale (1 = strongly disagree, 5 = strongly agree). Participants could end the exploration earlier if they wished, but each task lasted up to 15 minutes.

\paragraph{Post-Task questionnaire (5 minutes)} After exploration, participants completed a short Likert-scale questionnaire assessing the quality of the generated examples, the usefulness of \tool features, and overall feedback on their experience. The questions are shown in~\Cref{tab:survey_items}.

\paragraph{Interview (10 minutes)} Finally, we conducted semi-structured interviews to capture participants' subjective experiences on exploring distinguishing variation and how \tool might integrate into their existing development workflows.

\subsubsection{Materials} \label{ssec:studycomp}

\begin{figure*}
    \centering
    \includegraphics[width=1\linewidth]{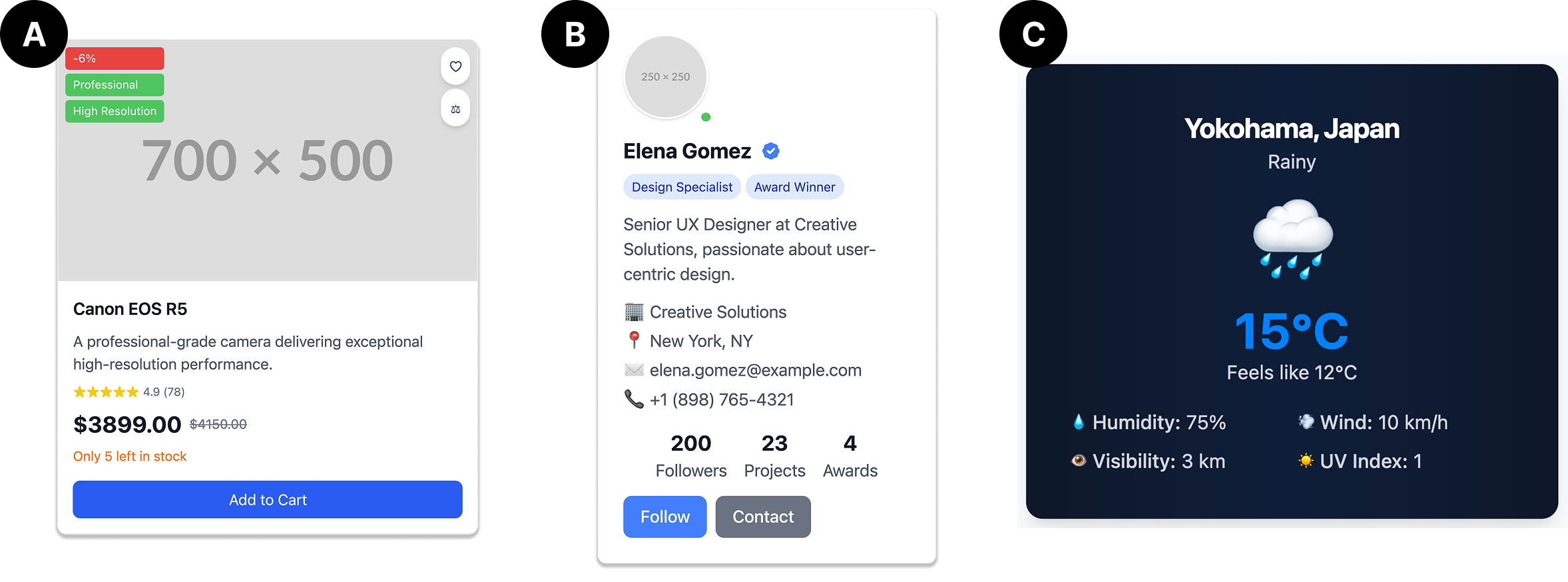}
    \caption{Screenshots of the three UI components used for user study. \textmd{\protect\ccircle{A} Product card, \protect\ccircle{B} Profile card, and \protect\ccircle{C} Weather card.} }
    \Description{This image contains screenshots of the three UI components used in the user study. (1) Product card, (2) Profile card, and (3) Weather card.}
    \label{fig:components}
\end{figure*}

For both the component tasks and the open-ended task, we prepared a set of three complex UI components: \smalltt{ProductCard}, \smalltt{WeatherCard}, \smalltt{ProfileCard} (shown in~\Cref{fig:components}). These components were adapted from production versions of \sbook workbooks used within our organization to share UI designs, and were directly inspired by those encountered in our formative study. Each component is an industry-deployed component that was modified to remove organizational dependencies that weren't relevant to the experiment. On average, the components exposed $10-12$ properties spanning diverse data types, including categorical values, free-form strings, numeric inputs, and nested data objects. For each participant, we chose two of the three UI components for the tasks. The order and selection of the components were counter-balanced.

We deliberately selected and modified components to ensure they would be unfamiliar to participants. This decision reflects a common development practice: front-end developers frequently rely on third-party component libraries that they did not author and whose internal implementation details are opaque. By situating participants in a similar “closed-box” scenario, we sought to capture the challenges and strategies that arise when working with unfamiliar components. This choice enhances the ecological validity of our study, aligning task performance more closely with real-world development contexts.

\subsubsection{Measurement and Analysis}

\begin{table}[t]
\centering
\small
\caption{Post-task questionnaire items to assess the variations. \textmd{(1 = Strongly Disagree, 5 = Strongly Agree).}\label{tab:survey_items}}
\begin{tabular}{p{\linewidth}}
\toprule
\textbf{Post-task questionnaire items (5-point Likert)} \\
\midrule
\begin{minipage}[t]{\linewidth}
\begin{enumerate}
\item I was able to explore a wide range of variations for this component.
\item The tool helped me understand all the different variations/possibilities for the component.
\item I was able to identify the most important properties of the component.
\item The examples felt natural and realistic.
\item I feel confident in applying this component in practice after exploration.
\item The coverage analysis was helpful in understanding missing examples.
\item The code editing was useful to understand components.
\end{enumerate}
\vspace{2pt}
\end{minipage}
\\
\bottomrule
\end{tabular}
\end{table}

\paragraph{Quantitative metrics}
During the exploration tasks, we logged participant interactions with \tool, including the number of variations generated, the sampling instructions, and in-place code edits. After each iteration, we also recorded participants' ratings of generated variations on three metrics: (1) natural, (2) distinct, and (3) coverage. After the two exploration tasks, participants completed short questionnaires using 5-point Likert scales with 1 representing \textit{Strongly Disagree} and 5 representing \textit{Strongly Agree} (\Cref{tab:survey_items}).

\paragraph{Qualitative interviews}
We conducted semi-structured interviews at the end of each session. Interviews were audio-recorded, transcribed, and thematically analyzed using an inductive coding approach. Themes were identified through collaborative discussion and supported by representative quotations.
We integrated findings across log data, post-task survey responses, and interviews to provide a holistic account of participants' engaged with \tool.

\section{User Study Results}

Overall, participants engaged actively with \tool during tasks, producing a diverse set of configurations, and expressed that the generated variations felt more natural and distinctive compared to their traditional workflows.

\subsection{Quantitative Results}

\begin{table}[h!]
\centering
\footnotesize
\caption{Likert Scale Ratings by Participant, Task, and Iteration. \textmd{Overall, participants rated variations as highly natural and realistic, indicating that the variations were mimetic. The ratings for distinct started high but trended downwards, whereas the ratings for coverage started low and trended upwards.}}
\label{tab:likert_heatmap}
\midsepremove
\resizebox{\columnwidth}{!}{
\begin{tabular}{llcccccccccccc}
\toprule
 &  & \multicolumn{4}{c}{\textbf{Natural and Realistic}} & \multicolumn{4}{c}{\textbf{Distinct}} & \multicolumn{4}{c}{\textbf{Coverage}} \\
\cmidrule(lr){3-6} \cmidrule(lr){7-10} \cmidrule(lr){11-14}
\textbf{PID} & \textbf{Task} & \textbf{I1} & \textbf{I2} & \textbf{I3} & \textbf{I4} & \textbf{I1} & \textbf{I2} & \textbf{I3} & \textbf{I4} & \textbf{I1} & \textbf{I2} & \textbf{I3} & \textbf{I4} \\
\midrule
P1 & T1 & \cellcolor{blue!40} $5$ & \cellcolor{blue!40} $5$ & \cellcolor{white} $\cdot$ & \cellcolor{white} $\cdot$ & \cellcolor{blue!10} $4$ & \cellcolor{blue!10} $4$ & \cellcolor{white} $\cdot$ & \cellcolor{white} $\cdot$ & \cellcolor{gray!10} $3$ & \cellcolor{blue!10} $4$ & \cellcolor{white} $\cdot$ & \cellcolor{white} $\cdot$ \\
P1 & T2 & \cellcolor{blue!40} $5$ & \cellcolor{blue!40} $5$ & \cellcolor{blue!40} $5$ & \cellcolor{white} $\cdot$ & \cellcolor{blue!10} $4$ & \cellcolor{blue!10} $4$ & \cellcolor{blue!40} $5$ & \cellcolor{white} $\cdot$ & \cellcolor{orange!10} $2$ & \cellcolor{blue!40} $5$ & \cellcolor{blue!40} $5$ & \cellcolor{white} $\cdot$ \\
P2 & T1 & \cellcolor{blue!40} $5$ & \cellcolor{blue!40} $5$ & \cellcolor{white} $\cdot$ & \cellcolor{white} $\cdot$ & \cellcolor{blue!10} $4$ & \cellcolor{blue!40} $5$ & \cellcolor{white} $\cdot$ & \cellcolor{white} $\cdot$ & \cellcolor{gray!10} $3$ & \cellcolor{blue!40} $5$ & \cellcolor{white} $\cdot$ & \cellcolor{white} $\cdot$ \\
P2 & T2 & \cellcolor{blue!40} $5$ & \cellcolor{blue!40} $5$ & \cellcolor{white} $\cdot$ & \cellcolor{white} $\cdot$ & \cellcolor{blue!40} $5$ & \cellcolor{blue!40} $5$ & \cellcolor{white} $\cdot$ & \cellcolor{white} $\cdot$ & \cellcolor{gray!10} $3$ & \cellcolor{blue!40} $5$ & \cellcolor{white} $\cdot$ & \cellcolor{white} $\cdot$ \\
P3 & T1 & \cellcolor{blue!40} $5$ & \cellcolor{blue!40} $5$ & \cellcolor{blue!10} $4$ & \cellcolor{white} $\cdot$ & \cellcolor{gray!10} $3$ & \cellcolor{gray!10} $3$ & \cellcolor{blue!10} $4$ & \cellcolor{white} $\cdot$ & \cellcolor{orange!10} $2$ & \cellcolor{orange!10} $2$ & \cellcolor{blue!10} $4$ & \cellcolor{white} $\cdot$ \\
P3 & T2 & \cellcolor{blue!40} $5$ & \cellcolor{blue!40} $5$ & \cellcolor{blue!40} $5$ & \cellcolor{white} $\cdot$ & \cellcolor{blue!40} $5$ & \cellcolor{blue!40} $5$ & \cellcolor{blue!10} $4$ & \cellcolor{white} $\cdot$ & \cellcolor{blue!40} $5$ & \cellcolor{blue!40} $5$ & \cellcolor{blue!40} $5$ & \cellcolor{white} $\cdot$ \\
P4 & T1 & \cellcolor{blue!10} $4$ & \cellcolor{blue!10} $4$ & \cellcolor{blue!40} $5$ & \cellcolor{white} $\cdot$ & \cellcolor{blue!10} $4$ & \cellcolor{blue!40} $5$ & \cellcolor{gray!10} $3$ & \cellcolor{white} $\cdot$ & \cellcolor{gray!10} $3$ & \cellcolor{blue!40} $5$ & \cellcolor{blue!40} $5$ & \cellcolor{white} $\cdot$ \\
P4 & T2 & \cellcolor{blue!40} $5$ & \cellcolor{blue!40} $5$ & \cellcolor{blue!40} $5$ & \cellcolor{white} $\cdot$ & \cellcolor{blue!40} $5$ & \cellcolor{gray!10} $3$ & \cellcolor{gray!10} $3$ & \cellcolor{white} $\cdot$ & \cellcolor{orange!10} $2$ & \cellcolor{blue!40} $5$ & \cellcolor{blue!40} $5$ & \cellcolor{white} $\cdot$ \\
P5 & T1 & \cellcolor{blue!10} $4$ & \cellcolor{blue!10} $4$ & \cellcolor{white} $\cdot$ & \cellcolor{white} $\cdot$ & \cellcolor{blue!40} $5$ & \cellcolor{gray!10} $3$ & \cellcolor{white} $\cdot$ & \cellcolor{white} $\cdot$ & \cellcolor{orange!40} $1$ & \cellcolor{blue!40} $5$ & \cellcolor{white} $\cdot$ & \cellcolor{white} $\cdot$ \\
P5 & T2 & \cellcolor{blue!10} $4$ & \cellcolor{blue!10} $4$ & \cellcolor{blue!10} $4$ & \cellcolor{white} $\cdot$ & \cellcolor{orange!10} $2$ & \cellcolor{gray!10} $3$ & \cellcolor{orange!10} $2$ & \cellcolor{white} $\cdot$ & \cellcolor{blue!10} $4$ & \cellcolor{blue!10} $4$ & \cellcolor{blue!40} $5$ & \cellcolor{white} $\cdot$ \\
P6 & T1 & \cellcolor{blue!10} $4$ & \cellcolor{blue!40} $5$ & \cellcolor{blue!10} $4$ & \cellcolor{white} $\cdot$ & \cellcolor{blue!40} $5$ & \cellcolor{blue!40} $5$ & \cellcolor{blue!10} $4$ & \cellcolor{white} $\cdot$ & \cellcolor{gray!10} $3$ & \cellcolor{blue!40} $5$ & \cellcolor{blue!40} $5$ & \cellcolor{white} $\cdot$ \\
P6 & T2 & \cellcolor{blue!40} $5$ & \cellcolor{blue!40} $5$ & \cellcolor{white} $\cdot$ & \cellcolor{white} $\cdot$ & \cellcolor{blue!40} $5$ & \cellcolor{blue!40} $5$ & \cellcolor{white} $\cdot$ & \cellcolor{white} $\cdot$ & \cellcolor{blue!10} $4$ & \cellcolor{blue!40} $5$ & \cellcolor{white} $\cdot$ & \cellcolor{white} $\cdot$ \\
P7 & T1 & \cellcolor{blue!10} $4$ & \cellcolor{blue!10} $4$ & \cellcolor{white} $\cdot$ & \cellcolor{white} $\cdot$ & \cellcolor{orange!10} $2$ & \cellcolor{gray!10} $3$ & \cellcolor{white} $\cdot$ & \cellcolor{white} $\cdot$ & \cellcolor{orange!40} $1$ & \cellcolor{blue!10} $4$ & \cellcolor{white} $\cdot$ & \cellcolor{white} $\cdot$ \\
P7 & T2 & \cellcolor{blue!10} $4$ & \cellcolor{blue!10} $4$ & \cellcolor{blue!10} $4$ & \cellcolor{white} $\cdot$ & \cellcolor{blue!40} $5$ & \cellcolor{blue!10} $4$ & \cellcolor{gray!10} $3$ & \cellcolor{white} $\cdot$ & \cellcolor{gray!10} $3$ & \cellcolor{blue!10} $4$ & \cellcolor{blue!40} $5$ & \cellcolor{white} $\cdot$ \\
P8 & T1 & \cellcolor{blue!40} $5$ & \cellcolor{blue!40} $5$ & \cellcolor{white} $\cdot$ & \cellcolor{white} $\cdot$ & \cellcolor{blue!40} $5$ & \cellcolor{blue!40} $5$ & \cellcolor{white} $\cdot$ & \cellcolor{white} $\cdot$ & \cellcolor{blue!40} $5$ & \cellcolor{blue!40} $5$ & \cellcolor{white} $\cdot$ & \cellcolor{white} $\cdot$ \\
P8 & T2 & \cellcolor{blue!40} $5$ & \cellcolor{blue!40} $5$ & \cellcolor{blue!40} $5$ & \cellcolor{white} $\cdot$ & \cellcolor{blue!40} $5$ & \cellcolor{blue!40} $5$ & \cellcolor{blue!40} $5$ & \cellcolor{white} $\cdot$ & \cellcolor{gray!10} $3$ & \cellcolor{blue!10} $4$ & \cellcolor{blue!40} $5$ & \cellcolor{white} $\cdot$ \\
P9 & T1 & \cellcolor{blue!10} $4$ & \cellcolor{blue!40} $5$ & \cellcolor{blue!10} $4$ & \cellcolor{blue!40} $5$ & \cellcolor{orange!10} $2$ & \cellcolor{blue!40} $5$ & \cellcolor{gray!10} $3$ & \cellcolor{blue!10} $4$ & \cellcolor{orange!10} $2$ & \cellcolor{blue!40} $5$ & \cellcolor{blue!10} $4$ & \cellcolor{gray!10} $3$ \\
P9 & T2 & \cellcolor{blue!10} $4$ & \cellcolor{blue!10} $4$ & \cellcolor{blue!10} $4$ & \cellcolor{blue!40} $5$ & \cellcolor{blue!40} $5$ & \cellcolor{blue!40} $5$ & \cellcolor{gray!10} $3$ & \cellcolor{blue!40} $5$ & \cellcolor{gray!10} $3$ & \cellcolor{blue!10} $4$ & \cellcolor{blue!10} $4$ & \cellcolor{blue!40} $5$ \\
P10 & T1 & \cellcolor{blue!10} $4$ & \cellcolor{blue!10} $4$ & \cellcolor{blue!10} $4$ & \cellcolor{blue!10} $4$ & \cellcolor{blue!40} $5$ & \cellcolor{blue!10} $4$ & \cellcolor{blue!10} $4$ & \cellcolor{orange!10} $2$ & \cellcolor{orange!10} $2$ & \cellcolor{blue!10} $4$ & \cellcolor{blue!10} $4$ & \cellcolor{blue!40} $5$ \\
P10 & T2 & \cellcolor{blue!10} $4$ & \cellcolor{blue!10} $4$ & \cellcolor{blue!10} $4$ & \cellcolor{white} $\cdot$ & \cellcolor{blue!40} $5$ & \cellcolor{blue!40} $5$ & \cellcolor{orange!40} $1$ & \cellcolor{white} $\cdot$ & \cellcolor{blue!40} $5$ & \cellcolor{blue!10} $4$ & \cellcolor{blue!10} $4$ & \cellcolor{white} $\cdot$ \\
P11 & T1 & \cellcolor{blue!40} $5$ & \cellcolor{blue!40} $5$ & \cellcolor{blue!10} $4$ & \cellcolor{white} $\cdot$ & \cellcolor{blue!10} $4$ & \cellcolor{blue!40} $5$ & \cellcolor{blue!10} $4$ & \cellcolor{white} $\cdot$ & \cellcolor{gray!10} $3$ & \cellcolor{blue!40} $5$ & \cellcolor{blue!40} $5$ & \cellcolor{white} $\cdot$ \\
P11 & T2 & \cellcolor{blue!40} $5$ & \cellcolor{blue!40} $5$ & \cellcolor{white} $\cdot$ & \cellcolor{white} $\cdot$ & \cellcolor{blue!40} $5$ & \cellcolor{blue!10} $4$ & \cellcolor{white} $\cdot$ & \cellcolor{white} $\cdot$ & \cellcolor{blue!40} $5$ & \cellcolor{blue!40} $5$ & \cellcolor{white} $\cdot$ & \cellcolor{white} $\cdot$ \\
P12 & T1 & \cellcolor{gray!10} $3$ & \cellcolor{blue!40} $5$ & \cellcolor{white} $\cdot$ & \cellcolor{white} $\cdot$ & \cellcolor{blue!40} $5$ & \cellcolor{blue!40} $5$ & \cellcolor{white} $\cdot$ & \cellcolor{white} $\cdot$ & \cellcolor{gray!10} $3$ & \cellcolor{blue!40} $5$ & \cellcolor{white} $\cdot$ & \cellcolor{white} $\cdot$ \\
P12 & T2 & \cellcolor{blue!40} $5$ & \cellcolor{blue!10} $4$ & \cellcolor{blue!40} $5$ & \cellcolor{white} $\cdot$ & \cellcolor{blue!10} $4$ & \cellcolor{orange!10} $2$ & \cellcolor{orange!10} $2$ & \cellcolor{white} $\cdot$ & \cellcolor{blue!10} $4$ & \cellcolor{blue!40} $5$ & \cellcolor{blue!40} $5$ & \cellcolor{white} $\cdot$ \\
\bottomrule
\end{tabular}
}

\vspace{2pt}
{\footnotesize
Likert responses:
\tikz \filldraw[color=orange!40] (0,0) rectangle (5pt,5pt); Strongly Disagree,
\tikz \filldraw[color=orange!10] (0,0) rectangle (5pt,5pt); Disagree,
\tikz \filldraw[color=gray!10] (0,0) rectangle (5pt,5pt); Neutral,
\tikz \filldraw[color=blue!10] (0,0) rectangle (5pt,5pt); Agree,
\tikz \filldraw[color=blue!40] (0,0) rectangle (5pt,5pt); Strongly Agree,
\tikz \filldraw[fill=white, draw=black] (0,0) rectangle (5pt,5pt); No Data.
}

\end{table}

Across tasks, participants explored on average $13.1$ variations ($min=9$, $max=20$, $\sigma = 3.2$) per component over an average of $2.87$ iterations ($min=2$, $max=4$, $\sigma = 0.67$). After every iteration, participants self-rated the generated variants on three key metrics: \emph{naturalness}, \emph{distinctiveness}, and \emph{coverage}, shown in~\Cref{tab:likert_heatmap}. Participants consistently rated the examples as highly natural and realistic across tasks and UI components. The rating for distinctiveness overall started high in the first iteration, but trended downward as more examples were generated. This decline suggests a natural saturation effect; as the system generates more examples, the probability of creating a truly novel or distinct variant decreases. In contrast, the participants' overall ratings for coverage started low but trended upward with each new set of examples. This indicates that participants felt the additional variants were effectively exploring and filling the design space, confirming that the system was successfully expanding the perceived boundaries of the design space with each iteration. We discuss this in more detail in the qualitative results.

\begin{figure*}[h]
    \centering
    \includegraphics[width=0.8\linewidth]{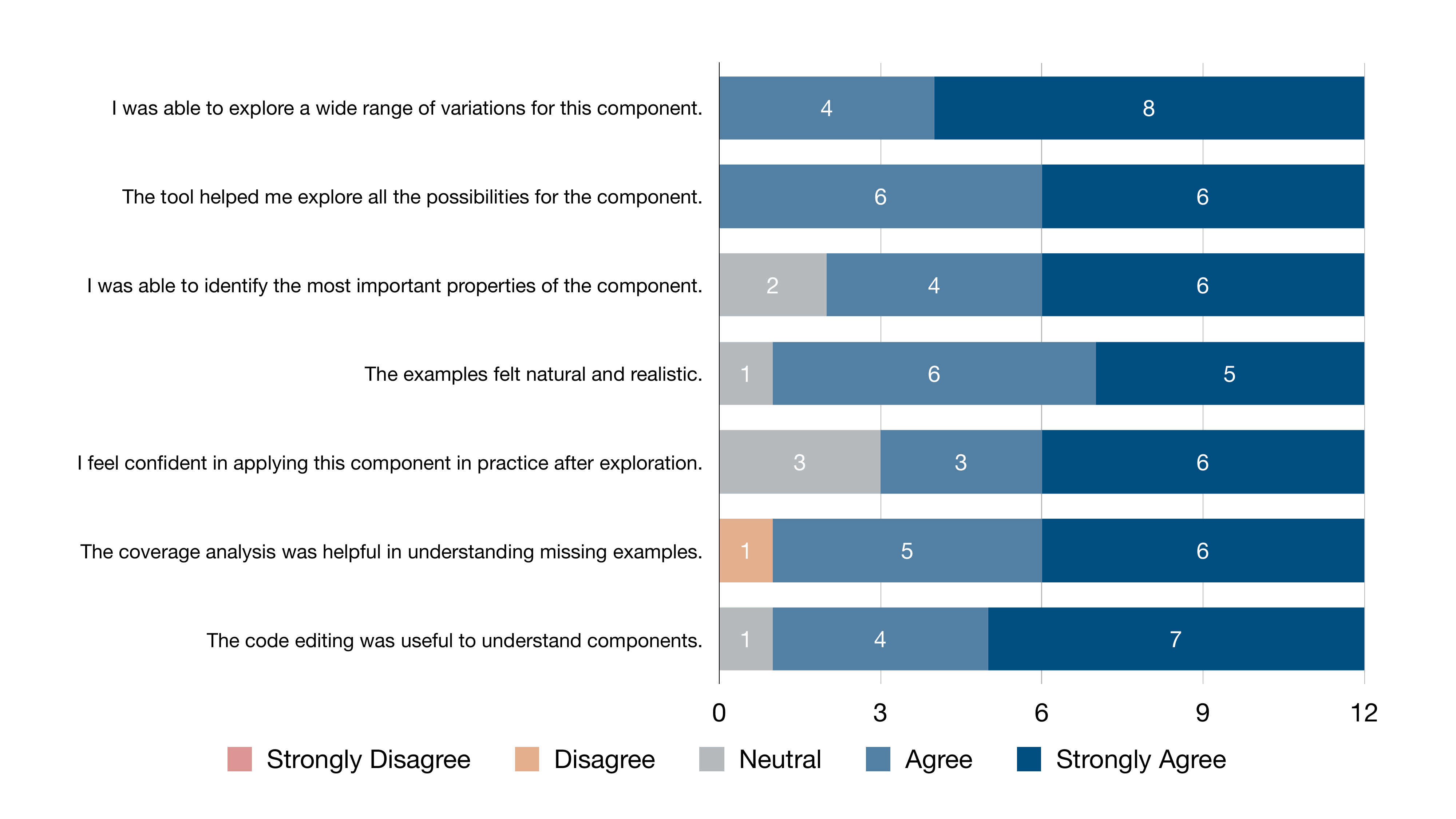}
    \caption{Post-task questionnaire results (5-point Likert scale). \textmd{Most participants agreed the variations were natural and realistic, explored all the possibilities of the component, and understood key component properties. They also rated \tool's coverage analysis and in-place code editing as useful.}}
    \Description{A chart showing post-task questionnaire results in a 5-point Likert scale. Most participants found the variations natural, could explore a wide range of examples, and understood key component properties. They also rated \tool's coverage analysis and in-place code editing as useful.}
    \label{fig:posttask}
\end{figure*}

\subsubsection{Post-task questionnaire responses} 

Post-task ratings (5-point Likert) indicated strong perceived support for design-space exploration (shown in~\Cref{fig:posttask}). All participants agreed ($rating>3$) that they could explore a broad range of variations (12/12) and that the tool helped participants cover the entire space of possibilities for each component (12/12). Most reported being able to identify the properties that mattered most (10/12), and examples were generally rated as natural and realistic (11/12). Despite encountering the UI components for the first time, 75\% of the participants were confident they could apply them in practice (9/12). 
Regarding specific system features, nine participants also agreed that \emph{Coverage Analysis} helped reveal missing regions and steering generation toward underrepresented areas. Finally, eleven participants agreed that the \emph{In-place code editor} in the inspect UI was useful to perform low-level edits to understand the inner workings of the component.

\subsection{Qualitative Results}

\subsubsection{Mimetic examples brought components to life}

All the participants consistently described the generated variations as natural and realistic, making \say{the examples just come to life} (P11) \fbox{RQ3}. Rather than abstract placeholder values that feel inanimate, the mimetic variations felt animated and grounded: \say{it's almost human-like examples and it tries to put all the right properties and all the relevant values} (P10). Participants were also delighted by how the tool seemed to understand the component and its domain. For the \emph{WeatherCard}, \say{it is already taking into account what temperature is reasonable for the location. It is showing Moscow for snowing and Cairo for sunny. It actually did a pretty good job of matching the location and the weather.} (P2). For the \emph{ProductCard}, \say{This shows me cards in different languages and a wide variety of layouts. It is even changing the currency based on the language. That is cool} (P9). 

Rendering with mimetic data makes the component resemble \say{things people would have encountered many times. So it is easy to point out if there is something unnatural or missing} (P12) \fbox{RQ3}. This contrasted with hand-written examples, which often \say{only render the examples that are important to us [developers], based on our requirements and constraints. But the users can have a different set of requirements, constraints, and what is important to them.} (P6) Using \tool, participants generated variations tailored to their own requirements.

\subsubsection{CoverageAnalysis as a map for exploration}

Participants described \emph{CoverageAnalysis} as more than just feedback---it became their compass for navigating the design space. As P3 explains, it provided a \say{comprehensive view of the possibilities rather than having to generate them one by one}. CoverageAnalysis became a map that revealed uncharted territories, waiting to be explored. This encouraged systematic exploration: \say{right off the bat I was able to realize that the colorful layout was not represented in the examples, and I could prompt it to see those variations. It helped me identify the gaps in the examples} (P6) \fbox{RQ1}.
Eleven participants noted that they also actively relied on \emph{CoverageAnalysis} to orient themselves: \say{I can just look at the coverage analysis, and match it with the examples to understand what props are affecting what in the components.} (P11) \fbox{RQ2}.

\subsubsection{Sampling instructions as gateways for serendipity}

While CoverageAnalysis promoted systematic mapping of design space, sampling instructions served as gateways that teleported participants to inspiring design spaces. Participants provided a wide variety of instructions to explore localization, interesting property combinations, and interface constraints. For instance, P12 instructed \tool to \say{generate profiles in 5 different languages} to examine how the component looks in other languages. In fact, six participants instructed the tool to explore variations in languages other than English, since many of their web pages were globally accessible in local languages. This was surprising, given that not a single UI component we inspected from the workbenches shared in our formative study contained variations in languages other than English. Four participants also gave instructions to combine properties in specific ways, such as \say{Create profile cards which only contain phone number and follower counts. Use very diverse names from different parts of the world} (P3) or \say{create a profile card for professional social media} (P9). Participants even grounded sampling on the visual aspects of the UI, like \say{show me the best examples for watch UI} (P8). Participants emphasized that these serendipitous variations revealed possibilities they hadn’t initially considered---like secret doors opening into new design directions. These are explorations that are never part of static documentation currently maintained in component libraries \fbox{RQ1}.

\subsubsection{Visual overview as a gallery of possibilities}

Existing documentation for component libraries often only \say{share the code examples, and it just wasn't as useful as we anticipated} (P3). 
Participants appreciated being able to wander visually through the design space rather than parsing lines of code. As one noted, \say{The visual comparison helps; otherwise, I have to go in and manually inspect all the different properties} (P7). Five participants highlighted the usefulness of a gallery-like view: \say{it's very visual instead of like relying on a lot of code examples or documentation} (P1) \fbox{RQ1}.

\subsubsection{Over-sampling as reassurance for trust}

While \emph{CoverageAnalysis} provided a comprehensive view of the property space and its coverage, six participants wanted to always see 
\say{just a few more examples} (P3). This desire was less about missing coverage and more about trust. They consider redundancy and repetition as a way of convincing themselves that no hidden surprises remain. 
As P12 explained, \say{the analysis says everything is covered, but I still think I need more examples to be confident I have seen all the options.} This also acts as a way of exploring variations that are outside the scope of just coverage, but can reveal interesting property choices: \say{even though everything's represented over here, at least once inside of these [examples], there's probably different variations that I might be interested in that's not shown here} (P7).

\subsubsection{In-place editor for playful tinkering}

The in-place editor gave participants a sense of play and control---what one called \say{a sandbox to play around} (P11). The edit-inspect loop became a way of hypothesis and discovery: \say{Looks like there is a border property that makes it look glass-y} (P4) \fbox{RQ2}.
The tinkering process generally started with selecting a variation, adjusting a property or two, and immediately watching the design transform. 
This edit–inspect cadence helped isolate which properties mattered visually and clarified cross-property interactions. P7 delightedly described this process: \say{for instance on this component, the avatar is over to the left, and if I just wanted to change the position, I can quickly just come over here [code-editor] and change it from left to the center, and so it's really kind of tweaking and adjusting a [variation] that I'm interested in.}

Four participants highlighted the value of moving seamlessly between the high-level visual overview (to spot interesting regions) and the low-level editor (to play around or refine a state). They described a workflow of \say{browse to find, edit to refine}: the visual overview helped them select a promising starting point; the editor gave them control to refine it. \say{Not only can I see it from a more high level point of view, but I can also make small code changes, and copy the code to use it in my code.} (P8) \fbox{RQ3}.

\subsubsection{Craving the edges: unnatural and adversarial cases}

While mimetic examples were praised as the right default for exploration, five participants wanted to push into the darker edges of the map---unnatural, adversarial, and extreme inputs \fbox{RQ4}.
As P12 put it, \say{But I also want to be able to break the component by generating unrealistic inputs.}
They asked for quick access to outliers: very long strings, missing fields, contradictory states. For some, this felt like testing the resilience of the component; for others, it was simply good practice. As P9 reflected, \say{having this dynamic generation with extreme values would help in testing the limits of a component after we implement it.} Combining such adversarial values with CoverageAnalysis, P5 suggested, would allow \say{generating many many examples for visual regression testing} \fbox{RQ4}. However, participants were often unsuccessful in generating unnatural and adversarial data, since the current implementation of the \naturalSampler was optimized to generate mimetic data.


\section{Discussion}

Our evaluation of \tool demonstrates that distinguishing variations successfully supports exploration and discovery of component design spaces. We now discuss the broader implications of our findings, considering how they inform tooling for UI development, the role of AI in developer workflows, and remaining challenges.

\paragraph{\say{The way we notice} is subjective}
When rating \emph{distinctness} between variations, participants differed in how they interpreted what counted as a meaningful change. Some were highly sensitive (and interested) to subtle visual artifacts such as shadows or borders, whereas others dismissed these as minor and instead wanted to \say{see big chunky changes} (P2). This suggests an opportunity for future research to explore adjustable scales of distinctness. Within \tool, such sensitivity could be tuned by modifying the visual impact scoring function (see~\Cref{sec:component-analyzer-module}).

\paragraph{Balancing breadth and focus in design-space exploration}
A central tension in exploring component design space is balancing exhaustiveness with clarity: too few variations obscure critical behaviors, while too many overwhelm. Our results suggest that \tool's hybrid strategy---symbolic pruning to highlight impactful properties, paired with LLM-based mimetic sampling---helps developers strike this balance. Participants not only encountered a broader range of examples compared to usual workflows, but also judged many to be meaningfully distinct rather than repetitive or trivial. This highlights design-space sampling as a valuable lens for supporting documentation, onboarding, and decision-making in component libraries.

\paragraph{Personalized sampling}
During the study, five participants wanted immediate access to \tool use with their own component libraries, particularly to generate variations grounded in domain-specific data. As P6 explained, \say{We keep the components and the stories free of application logic and generic. This means it becomes very hard to track how it might actually look in the application. This is where [\tool{}] can be helpful. You can just mock data structures mapped to the applications to really help target the story we want.} Participants noted that this ability to bring in realistic, application-relevant data would bridge the gap between generic component libraries and actual use cases, helping them anticipate how components might perform in production. Several participants remarked that such domain-sensitive examples would not only improve confidence in their designs but also support communication with non-technical stakeholders, like designers and product managers, by showing the components in context.

\paragraph{Beyond LLM-based samplers.}
While our \naturalSampler~leveraged the world knowledge of large language models to generate realistic values, other forms of samplers could also be used to produce distinguishing variations. Domain-specific samplers, for instance, could draw from structured datasets (e.g., weather archives, e-commerce catalogs, or social media profiles) to yield grounded examples tied directly to a target application. Generative models such as diffusion or GAN-based image/text generators could enrich the visual or linguistic diversity of component content, pushing examples toward more imaginative or stylistic extremes. Symbolic or constraint-based samplers combined with an archive of domain data (or heuristics) could ensure systematic coverage of rare or adversarial cases, complementing mimetic examples with robustness testing. Exploring hybrid pipelines---where multiple samplers are orchestrated depending on developer intent (e.g., mimetic realism, adversarial stress testing, or stylistic exploration)---could further expand the expressive power of design-space sampling.

\paragraph{Design Implications}

We highlight three design implications for future tools:  
\begin{enumerate}
    \item \textbf{Integrated exploration and transfer.} Participants wished for tighter coupling between explored variations and production code. Embedding mechanisms to export or \say{commit} selected configurations would make the tool more actionable.  
    \item \textbf{Guided contrastive exploration.} Participants valued seeing variations side-by-side. Future systems might explicitly scaffold contrastive workflows (e.g., \say{show me differences across a selected set of variations}) to better surface the trade-offs developers need to reason about.  
    \item \textbf{Trust through transparency.} While participants appreciated the \emph{natural-ness} of AI output, transparency about how variations were sampled, and the ability to adjust parameters, would help maintain trust and allow developers to calibrate exploration according to context.  
\end{enumerate}

\paragraph{Limitations and future work}
Due to the scope of \tool and the constraints of \sbook, we currently focus only on \emph{visual} differences among variations, and not on \emph{behavioral} differences. To understand component usage in practice, both matter, and we hope future work addresses this limitation. Further, our implementation of \naturalSampler uses LLMs in the backend, which are known to produce hallucinations and spurious outputs. We can identify syntax and type errors in the generated outputs and rectify them, but we still cannot reliably detect other extraneous outputs. This remains an open problem.

Our evaluation has several limitations. First, our participant pool ($12$) was relatively small and drawn from a single organization, which may limit generalizability to other development contexts. Second, our study components were adapted rather than deployed in production settings; while this design choice enhanced ecological validity, actual adoption may reveal additional needs such as collaboration support across teams. Third, while our hybrid approach reduces implausible generations, occasional errors indicate the need for stronger data validation and domain adaptation.

Future work could expand evaluation to larger and more diverse developer populations, investigate integration with continuous integration pipelines, and explore extensions to multi-component workflows where interactions between components compound design-space complexity. There is also an opportunity to adapt naturalistic sampling to other modalities (e.g., mobile UI, multimodal interaction widgets), broadening applicability beyond web-based component libraries.

\section{Conclusion}


In this paper, we proposed the concept of \emph{distinguishing variations}, which blends mimesis and distinctness to help developers explore the component design space and grasp how components behave across realistic yet meaningfully different scenarios. To generate such variations, we introduced a hybrid approach that integrates symbolic techniques with a mimetic sampler, and operationalized this approach in \tool. Our evaluation of \tool with professional developers demonstrated that distinguishing variations helped participants successfully discern and map the component design space, uncover serendipitous possibilities, and discover domain-relevant mimetic instantiations. From somewhere beyond the clouds, \tool invites us to notice that we are designed to love and break, and to rinse and repeat it all again~\cite{sheeran2014ed}.



\bibliographystyle{ACM-Reference-Format}
\bibliography{references}

\appendix


\section{\texorpdfstring{\naturalSampler~\texttt{Prompt}}{MimeticSampler Prompt}}\label{ap:prompt}

\begin{lstlisting}[language={},breaklines=true,breakatwhitespace=true,columns=fullflexible,
                   basicstyle=\ttfamily\footnotesize,keepspaces=true,showstringspaces=false]
You are an expert in statistical sampling and UI component design space exploration. Your role is to sample diverse, meaningful instantiations from a component's property configuration space. Given a structured component analysis describing available properties, their types, constraints, and usage code examples, you generate property value combinations that maximize coverage of the component's design space while maintaining semantic coherence and real-world plausibility.

**Component Property Space:**

Component: {componentName}
Has Children: {boolean}

*High Visual Impact Properties:*
For each property:
- Property name (dataType)
- Required: boolean
- Default value: if applicable
- Allowed values: for categorical properties
- Usage Examples: usage context / code examples

*Medium Visual Impact Properties:*
For each property:
- Property name (dataType)
- Required: boolean
- Default value: if applicable
- Allowed values: for categorical properties
- Usage Examples: usage context / code examples

*Low Visual Impact Properties:*
For each property:
- Property name (dataType)
- Required: boolean
- Default value: if applicable
- Allowed values: for categorical properties
- Usage Examples: usage context / code examples

**Sampling Strategy:**

### 1. Property Space Characterization
Before sampling, characterize the component's property space:
- **Dimensionality**: Identify all configurable properties (dimensions)
- **Value Domains**: For each property, determine the valid value space:
  * Categorical: Use provided `allowedValues`
  * Boolean: Sample {true, false}
  * Numeric: Sample boundary values, typical values, and edge cases
  * String: Generate domain-appropriate realistic values
  * Object/Array: Sample structurally diverse instances following the schema in `description`
- **Visual Impact Hierarchy**: Prioritize sampling high-impact properties more extensively than low-impact ones
- **Dependency Structure**: Identify property interactions where certain combinations are semantically invalid or particularly meaningful

### 2. Sampling Objectives
Generate N property configurations (where N = {story_count}) that achieve:
- Span the property space to represent diverse regions
- For high-impact properties: Sample all significant values
- For categorical properties: Include representative examples from each category
- For continuous properties: Sample boundaries, midpoints, and extrema
- Ensure sampled configurations represent realistic, meaningful use cases
- Generate domain-appropriate content
- Include property configurations representing canonical scenarios:
  - **Baseline**: Default/minimal configuration demonstrating core functionality
  - **Layout Variations**: Configurations testing responsive behavior, overflow handling, nested structures
  - **Visual Variants**: If the component has visual variant properties (size, color, style), sample across the variant space

### 3. Data Authenticity Requirements
When sampling string, object, or array property values:
- Generate **realistic, domain-appropriate data** based on real-world knowledge
- **Avoid** generic placeholders ("Lorem ipsum", "Example 1", "Test data", "User 1")
- **Match** appropriate semantic context for Geographic data, Names, Temporal data, Numeric data, E-commerce, Social content, etc. 
- For list-based properties (arrays, tables): Generate sufficient volume (5-10+ items) to demonstrate realistic usage patterns
- Follow user instructions about data requirements -- this is **very important**

### 4. Output Format
For each sampled configuration, generate a JSON object that instantiates all property values:

**Configuration Structure** (repeat for each of N samples):
```json
{
  "name": "DescriptiveName",
  "description": "Brief description of what this configuration demonstrates",
  "properties": {
    "propertyName1": "sampledValue1",
    "propertyName2": "sampledValue2",
    "propertyName3": ["array", "values", "if", "applicable"],
    "propertyName4": {
      "nested": "object",
      "if": "applicable"
    }
  }
}
```

### 5. Technical Conventions
- For image properties: Use `https://placehold.co/{width}x{height}` (e.g., `https://placehold.co/600x400`)
- For function properties: Describe the function's purpose as a string (e.g., "handleClick", "onSubmit")
- For complex object/array properties: Generate complete, realistic instances following the schema described in the property description

**Coverage Requirements**
{instructions_from_coverage_analyzer}

**User Instructions**
{custom_instructions}
\end{lstlisting}

\end{document}